\PassOptionsToPackage{hidelinks}{hyperref}
\documentclass[%
 reprint,
 superscriptaddress,
 amsmath,amssymb,
 aps,
 pra,
]{revtex4-1}

\usepackage{graphicx}
\usepackage{dcolumn}
\usepackage{bm}
\usepackage{hyperref}
\usepackage{longtable}
\usepackage[normalem]{ulem}
\usepackage[hidelinks]{hyperref}
\usepackage[T1]{fontenc}
\usepackage[utf8]{inputenc}
\usepackage{textcomp}
\usepackage{siunitx} 
\usepackage{booktabs}
\usepackage{siunitx}
\usepackage{graphicx} 
\usepackage{float}  
\usepackage{hyperref}
\hypersetup{
    colorlinks=true,
    linkcolor=blue,
    urlcolor=blue,  
}

\usepackage{xcolor}
\usepackage{amsfonts, amsmath, amssymb, amsthm}
\usepackage{braket}

\graphicspath{ {./figures/} }

\begin{document}

\preprint{APS/123-QED}

\title{Multi-Mode Quantum Memories for High-Throughput Satellite Entanglement Distribution}

\author{Connor Casey}
\email{cacasey@umass.edu}
\affiliation{Department of Physics, University of Massachusetts Amherst, Amherst, MA, USA}

\author{Albert Williams}
\affiliation{College of Information and Computer Sciences, University of Massachusetts Amherst, Amherst, MA, USA}

\author{Catherine McCaffrey}
\affiliation{Department of Physics, University of Massachusetts Amherst, Amherst, MA, USA}

\author{Eugene Rotherham}
\affiliation{Department of Space and Climate Physics, University College London, London, United Kingdom, London, United Kingdom}

\author{Nathan Darby}
\affiliation{Department of Physics, University of California, Santa Barbara, CA, USA}

\date{\today}

\begin{abstract}
Quantum networking seeks to enable global entanglement distribution through terrestrial and free-space channels; however, the exponential loss in these channels necessitates the need for quantum repeaters with efficient, long-lived quantum memories (QMs). Space-based architectures, particularly satellite-assisted links, offer a path to truly global connectivity; yet, they demand QMs that are compatible with orbital factors such as infrared radiation and the unique challenges of operating aboard a satellite. In this work, we propose a multimode quantum memory (MMQM) for Low-Earth Orbit (LEO) repeaters based on the atomic frequency comb (AFC) protocol. Our design integrates a hybrid alkali–noble-gas ensemble in an optical cavity, using alkali atoms for strong photon–matter coupling and noble-gas nuclear spins for minutes-to-hours coherence all without the need for cryogenics. The architecture natively supports temporal and spectral multiplexing, enabling the storage of 100 modes to parallelize probabilistic operations and overcome light-limited round-trip times. Representative link budgets at $h=500$ km with realistic apertures, $\eta_{\text{mem}}\!\gtrsim\!70\%$, and $t_{\text{buffer}}$ of several minutes predict improvements of up to two orders of magnitude in per-pass success probability and instantaneous SKR relative to a memoryless dual downlink, with clear scaling in $N$. Our contributions are: (i) a non-cryogenic, space-ready multimode memory; (ii) a systems analysis coupling mode count, storage time, and orbital geometry to achievable rate; and (iii) a near-term implementation roadmap. Together, these results indicate feasibility with current to near-term technology and provide a practical path toward a high-rate, space-enabled quantum internet.
\end{abstract}

\maketitle

\section{Introduction}
\label{sec:intro}
Quantum networking aims to interconnect quantum computers, sensors, and communication systems by distributing quantum states and entanglement across geographically separated nodes \cite{Knaut2024}. While metropolitan-scale demonstrations have validated key primitives, extending such capabilities to continental and ultimately global distances remains challenging due to loss, noise, and the fundamental constraints of quantum mechanics \cite{Aspelmeyer2018,Briegel1998}. Unlike classical signals, quantum states cannot be amplified to compensate for attenuation due to the no-cloning theorem, which states that there is no unitary operation that can duplicate an arbitrary quantum state \cite{nc}. Consequently, error mitigation strategies such as quantum error correction and entanglement purification must be engineered to tolerate loss without violating the linearity of quantum mechanics \cite{nc, VanMeter2014}.

Terrestrial channels can illustrate what these concepts mean for practical applications. In optical fibers, transmission losses grow exponentially with distance, which limits practical direct links to a few hundred kilometers, even under favorable conditions \cite{martin2021quantum,deforges2023satellite}. Free-space optical (FSO) links can mitigate exponential fiber loss by taking advantage of the inverse-square scaling of free-space channels; however, turbulence, weather, and line-of-sight requirements are additional sources of error that collectively work to degrade performance in these channels \cite{Aspelmeyer2018,rotherham2024advancing,vinay2017practical,sit2017high}. Satellite-based quantum communication leverages the near-vacuum environment of space and the favorable geometric scaling of free-space paths to enable intercontinental entanglement distribution, thus connecting isolated terrestrial clusters.  \cite{Pirandola2021,khatri2021spooky,deForgesdeParny2023}. In this hybrid architecture, satellites function as entanglement distribution hubs, feeding ground stations that, in turn, interface with optical fiber and local networks \cite{deForgesdeParny2023}.

Even with satellite links, end-to-end performance at global scales ultimately relies on quantum repeaters \cite{Briegel1998,Kalachev2023QuantumRepeaters}. Repeaters divide a long path into shorter elementary links, generate entanglement locally, and extend it across the network via a method known as entanglement swapping \cite{zangi2023entanglement}. For channels with high loss, techniques such as entanglement purification can ensure that the fidelity of the signal is maintained while creating end-to-end entanglement \cite{childs2023streaming}. Because these steps are inherently probabilistic, a hardware component known as quantum memory (QM) acts as a buffer to store successful entangled pairs while other attempts proceed. This, in turn, avoids exponentially small coincidence probabilities and restores the polynomial scaling of the overall success rate \cite{Briegel1998}. Experimental implementations of these devices, therefore, require interfaces that can absorb and re-emit qubits on demand with high efficiency and fidelity over application-appropriate storage times.

The requirements for space-based QMs are especially stringent when compared to their terrestrial counterparts. Many space scenarios demand adjustable delays from milliseconds to seconds—e.g., geostationary orbit to Earth ($\sim$120 ms (one way)) or Earth–Moon links ($\sim$1.3s)—to synchronize heralding signals and enable feed-forward across long baselines \cite{gundogan2021quantum}. Laboratory demonstrations have already achieved long-lived storage with high efficiency (e.g., $\eta>70\%$ at $\tau\sim200$ ms), illustrating feasibility in principle \cite{yang2016efficient}. By contrast, realizing comparable delays with fiber spools would incur prohibitive losses (e.g., $>5,600$ dB for hundreds of milliseconds), underscoring the necessity of true quantum memories rather than passive delay lines \cite{gundogan2021quantum}. Theory further indicates that space-deployed QMs can unlock truly global entanglement distribution by coordinating probabilistic operations over planetary distances \cite{liorni2021quantum,gundogan2021quantum}.

Multiple physical platforms are under active investigation for QMs, including ensemble-based media (e.g., cold and warm atomic gasses and rare-earth–ion–doped solids, REIDs) and single-emitter systems (e.g., NV centers, single atoms, and molecules). Optically active systems are particularly attractive because they natively interface with photonic qubits used for long-distance distribution. In repeater settings, these memories must support on-demand retrieval, high efficiency and fidelity, long coherence times commensurate with link latencies, and compatibility with entanglement swapping and purification. Cavity-enhanced atom–photon interactions (as in DLCZ-type schemes) provide one route to strong coupling for the generation and manipulation of stored excitations \cite{DLCZ}, while solid-state ensembles offer promising paths to integration and scalability.

Beyond these baseline requirements, multimode capacity emerges as a decisive lever for throughput. Modern networks achieve high data rates by exploiting parallelism; analogously, quantum networks benefit from memories that can store and process many independent photonic modes across temporal, spectral, or spatial degrees of freedom (or their combinations). As emphasized in prior work, “multimode capacity of [a] communication channel is an essential requirement for high data rates,” and a central challenge is realizing a quantum memory that simultaneously accommodates multiple single-photon modes \cite{nature1650}. The performance impact is immediate at the link level: if two single-mode memories are bridged by a channel of length $L_0$ and refractive index $n$, the trial rate for heralded entanglement is bounded by the communication time $\tau_{\mathrm{comm}}=nL_0/c$, yielding a maximum repetition rate $R=1/\tau_{\mathrm{comm}}$ \cite{simon2007quantum,ortu2022multimode}. For $L_0=100$ km of fiber, this limits trials to $R\approx2$ kHz. A memory with $N$ modes can parallelize attempts during each $\tau_{\mathrm{comm}}$, boosting the heralding rate by $\mathcal{O}(N)$ to first order \cite{simon2007quantum,ortu2022multimode}. In large-scale networks—particularly those incorporating satellite segments where round-trip times are long—such parallelism is necessary to achieve high-throughput entanglement distribution that meets the demands of practical applications. 

In this work, we propose a multimode quantum memory (MMQM) protocol based on the atomic frequency comb (AFC) protocol tailored to Low-Earth Orbit (LEO) satellite networks. Our architecture uses a hybrid alkali–noble-gas ensemble inside an optical cavity, which combines the strong optical interface of alkali atoms with the long-lived coherence of noble-gas nuclear spins. This hybrid approach offers an intrinsically multimode interface with access to temporal and spectral multiplexing, while also enabling coherence times ranging from minutes to hours. Our objective is to specify a space-compatible MMQM architecture that aligns with the unique constraints of satellites and the operating environment aboard a satellite. By integrating multimode storage, future satellite-repeater constellations could utilize operations such as entanglement swapping and purification to help realize a truly global quantum internet.

\begin{figure}[]
\centering
\includegraphics[width=\linewidth]{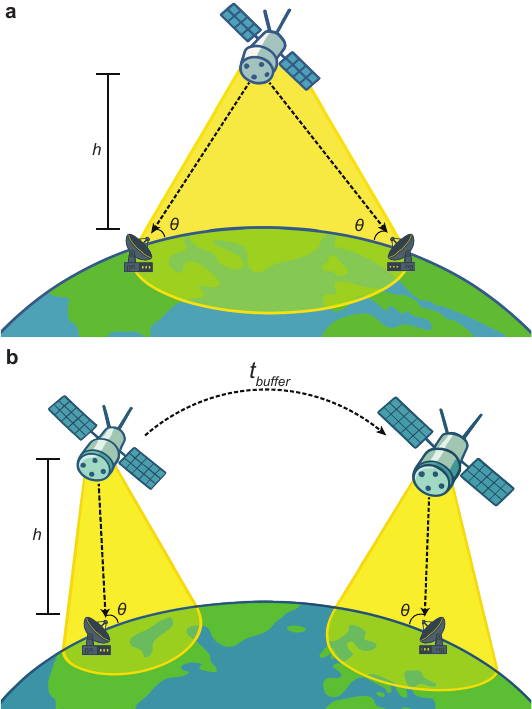}
\caption{\textbf{Buffered vs.\ dual downlink architectures.}
\textbf{(a) Dual downlink:} An entangled-photon source transmits to two ground stations only during their co-view window.
\textbf{(b) Buffered downlink (AFC-enabled):} One photon is downlinked while its partner is stored in an onboard AFC spin-wave memory; after a delay \(t_{\text{buffer}}\) (time between passes) the photon is recalled and sent to the second site, removing the simultaneous-visibility requirement. The memory must satisfy \(T_{\text{mem}}\!\ge\! t_{\text{buffer}}\) and contributes efficiency \(\eta_{\text{mem}}\) to the link budget. In both panels, \(h\) is satellite altitude, \(\theta\) the elevation angle; dashed lines are quantum channels and yellow cones are the satellite footprint.}
\label{fig:dual-vs-buffered}
\end{figure}

\section{Background}
\label{sec:background}

\subsection{Satellites, bosonic channels, and repeater motivation.}
Free-space optical (FSO) satellite links are modeled as lossy bosonic channels with transmissivity \(\eta\) set by diffraction, pointing jitter, and atmospheric effects. In the absence of quantum repeaters, the private and quantum capacities of such channels obey repeaterless bounds that degrade strongly with loss \cite{Pirandola2017FundamentalLimits,Pirandola2021FSOBounds}. Quantum repeaters mitigate this by distributing entanglement over shorter elementary links and performing entanglement swapping, at the cost of requiring QMs, which preserve states through probabilistic heralding and classical signaling latency \cite{Azuma2023RMP}. Critically, the required memory time scales inversely with the photonic transmission probability, \(T_{\text{mem}}\!\propto\!1/P_{\text{trans}}\) \cite{Zheng2022PRXQ040319}. Consequently, space-relevant memories should be (i) long-lived, (ii) optically interfaced for efficient photon–matter conversion, (iii) compatible with satellite constraints, and (iv) designed to function within the unique environment of space. 

\subsection{Link geometries: uplink, downlink, and dual downlink.}
In an uplink memory-assisted architecture, two ground stations send single photons (e.g., BB84-encoded) to a satellite that buffers them in distinct QMs; upon heralded loading of both memories, the satellite retrieves and interferes the photons in a Bell-state measurement (BSM) \cite{Abruzzo2014PRA,Panayi2014NJP}. Direct (QND) and indirect (entanglement-heralded) loading strategies have been analyzed \cite{Panayi2014NJP}. The principal advantage of this approach is a high repetition rate; however, early-path atmospheric diffraction induces additional loss from the "shower curtain," which is \(\sim\!10{-}20\) dB worse than downlink for comparable apertures. This shifts the burden to larger receivers in space and ultimately limits the range and key rate \cite{Gundogan2021npjQI}. Representative budgets (e.g., \(r_\text{Tx}\!=\!15\) cm ground, \(r_\text{Rx}\!=\!50\) cm space; \(\theta\!\sim\!10~\mu\)rad; 5 ms storage; \(80\%\) write–read; 20 MHz clock) show speedups over memoryless links to \(\sim\)1450 km but no straightforward path to multi-hop repeater nesting because photons propagate from ground\(\rightarrow\)space \cite{Gundogan2021npjQI}.

In a downlink variant, a central node emits photons entangled with each memory and transmits them to the ground stations. The central BSM then waits for classical heralds from the receivers, imposing a speed-of-light latency and hence second-scale storage times for \(\mathcal{O}(10^3)\) km baselines, unless parallelized via \(m\) memory pairs \cite{Luong2016APB,Gundogan2021npjQI}. 

Finally, dual-downlink entanglement distribution (E91) places the source onboard the satellite and sends one photon of each pair to each ground station. Because the satellite does not need to be trusted, this approach aligns well with security goals and with modern (biased-basis) QKD post-processing \cite{Bae2025arXiv,BennettBrassard2014TCS,Ekert1991PRL}. In all cases, the physical link is a bosonic channel, so performance hinges on combating loss and background noise while maximizing spatial/temporal/spectral multiplexing to make up for the probabilistic nature of these approaches.

\subsection{Space-induced decoherence and satellite resource limits.}
The environment of space differs dramatically from that of the laboratory. Natural ionizing radiation (cosmic rays, secondary particles) produces correlated errors and excess quasiparticles in solid-state devices, which degrades coherence and stability \cite{Mariani2023CosmicRaysQubits}. While photonic qubits are comparatively resilient, radiation and background light still load detectors and reduce state fidelity \cite{Fowler2024PRXQ}. Ultra-high-fidelity operations must also contend with vacuum-fluctuation effects (e.g., Lamb-shift-scale level shifts) and tight thermal budgets given passive radiative cooling \cite{Fragner2008Science}. From a systems engineering perspective, power, mass/volume, and heat-lift are  also scarce onboard a satellite, which constrains cryogenics, shielding, and onboard computing for real-time feed-forward/QEC \cite{Xing2024arXivCOTS}. Practical satellite nodes, therefore, favor compact, power-lean, and thermally tolerant memory platforms and detectors, such as space-qualified superconducting nanowire single-photon detectors (SNSPDs) \cite{You2018Opex}. Taken together, these environmental and resource constraints show that space-based hardware is more error prone and resource limited when juxtaposed with terrestrial laboratory systems. 

\subsection{Ensemble based quantum memory candidates for space.}
Ensemble based systems are especially attractive for space deployment because they interface directly with flying photons and entangled sources, support large optical bandwidths, and offer intrinsic multimode capacity in temporal, spectral, and spatial degrees of freedom \cite{Gundogan2021npjQI}. Platforms currently under investigation include the following: 

\begin{itemize}
    \item \textbf{Warm vapors} (alkali) operate without laser cooling or cryogenics and already show noiseless and fast operation, $\sim$GHz bandwidth, $\eta\!>\!80\%$ with fidelity above the no cloning bound and $\sim$1 s storage in SERF like regimes; coupling to noble gas nuclear spins via spin exchange offers a path to hour scale coherence \cite{Cho2016Optica,Dudin2013PRA,Katz2021SciAdv,Katz2022PRA}. Compact vapor cell systems also have space heritage as optical frequency references on rockets and nanosats \cite{Strangfeld2021JOSAB}.
    \item \textbf{Laser cooled atomic ensembles} (including Bose Einstein condensates (BECs)) provide high optical depth and low diffusion, enabling high efficiency, minute scale storage, and both temporal and spatial multiplexing \cite{Cho2016Optica,Dudin2013PRA,Heller2020PRL,Lan2009Opex,Pu2017NatComm}. Extensive microgravity and flight demonstrations together with experiments on the International Space Station attest to the growing attractiveness of this segment \cite{Langlois2018PRApplied,Barrett2016NatComm,Muntinga2013PRL,Deppner2021PRL,Becker2018Nature,Lachmann2021NatComm,Liu2018NatComm,Aveline2020Nature,Laurent2015CRPhys,Frye2021EPJQT,Devani2020CEAS,ElNeaj2020EPJQT}.
    \item \textbf{Rare earth ion doped crystals (REIDs)} at cryogenic temperatures combine narrow optical and spin transitions (long coherence) with strong multiplexing capability. Demonstrated performances include heralded intermemory entanglement, bright pulse storage from minutes to hours, and temporal, spectral, and spatial multimodality; integration via waveguides and nanophotonics improves compactness and light matter coupling \cite{LagoRivera2021Nature,Liu2021Nature,Heinze2013PRL,Usmani2012NatPhoton,Gundogan2013NJP,Seri2019PRL,Sinclair2014PRL,Saglamyurek2016NatComm,Yang2018NatComm,Saglamyurek2011Nature,Marzban2015PRL,Corrielli2016PRApplied,Zhong2017Science,Dibos2018PRL,Jobez2015PRL,Gundogan2015PRL,Businger2020PRL}. Hybridized electron and nuclear spin levels in anisotropic hosts are particularly attractive for long lived, on demand storage while maintaining large optical bandwidths \cite{Businger2020PRL}.
\end{itemize}

\subsection{Atomic Frequency Comb (AFC) and multimode scaling.}
AFC memories spectrally tailor an inhomogeneously broadened \(|g\rangle\!\leftrightarrow\!|e\rangle\) transition into a comb with tooth spacing \(\Delta\). An incident photon excites a collective Dicke state that rephases after \(2\pi/\Delta\), which yields an echo. A control pulse then maps this optical excitation to a spin state \(|s\rangle\) for on-demand retrieval and long-term storage. Backward retrieval can approach unit efficiency at high optical depth \cite{Afzelius2009PRA}. Importantly for repeaters, the AFC temporal multimode capacity is set by the number of teeth (i.e., the product of AFC bandwidth per channel, storage time, and the number of parallel channels) rather than by optical depth, as in EIT/Raman protocols \cite{Afzelius2009PRA}. Recent demonstrations at telecom wavelengths report the storage of \(\sim\)10\textsuperscript{3}–10\textsuperscript{4} modes (e.g., 1650 modes) with clear headroom via finer spectral packing and parallel spectral/spatial multiplexing \cite{Wei2024npjQI,Sinclair2014PRL,Seri2019PRL,Yang2018NatComm}. 

\subsection{Implications for a satellite multimode memory architecture.}
Coupling the bosonic-channel loss profile of FSO links and the inverse scaling of required memory time with transmission probability \cite{Zheng2022PRXQ040319} to the practicalities of space operation \cite{Fowler2024PRXQ,Fragner2008Science,Xing2024arXivCOTS}, AFC-based multimode ensemble memories emerge as natural building blocks for space repeaters. They (i) raise the per-pass success probability via large temporal/spectral mode counts, (ii) leverage spin (including nuclear) degrees of freedom for long coherence times, and (iii) admit power-lean hardware components that are compatible with satellite constraints \cite{Gundogan2021npjQI}.

\section{Protocol Design}
\label{sec:protocol-design}
\begin{figure}[h]
\centering
\includegraphics[width=\linewidth]{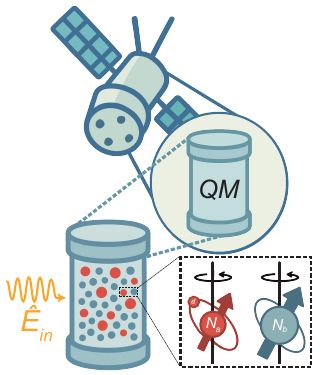}
\caption{\textbf{Satellite-borne hybrid alkali–noble-gas quantum memory.}
An ensemble of \(N_a\) alkali-metal atoms (red) and \(N_b\) noble-gas atoms (blue) is contained in a sealed glass cell mounted inside a single-ended optical cavity with one partially transmitting mirror. Prior to operation, the alkali spins are polarized along \(\hat{z}\) by optical pumping and the noble-gas spins are polarized via spin-exchange optical pumping (SEOP). The incident optical mode \(\hat{E}_{\mathrm{in}}\) drives the alkali transition \(|g\rangle\!\leftrightarrow\!|e\rangle\); the collective alkali spin \(\hat{S}\) couples to the noble-gas spin \(\hat{K}\) through spin exchange at rate \(J\), mapping the excitation to the long-lived nuclear-spin mode and allowing re-emission as \(\hat{E}_{\mathrm{out}}\).}
\label{fig:sat_mem}
\end{figure}

As seen in Figure \ref{fig:sat_mem}, we consider an ensemble of $N_a$ alkali-metal and $N_b$ noble-gas atoms contained within a sealed glass cell mounted inside an optical cavity with one partially transmitting mirror, following the architecture of the original terrestrial design introduced in \cite{foundationalPaper,ogCRYOpaper}. Initially, the alkali spins are polarized along $\hat{z}$ via optical pumping, while the noble-gas spins are polarized through spin-exchange optical pumping techniques. 

The alkali atoms are modeled as $\Lambda$-systems with a ground state $\ket{g}$ and a metastable state $\ket{s}$ coupled optically to the excited state $\ket{e}$, while noble-gas atoms are treated as two-level systems $\ket{\Downarrow}$ and $\ket{\Uparrow}$. The quantum signal field $\hat{E}$ couples $\ket{g}$ to $\ket{e}$, and a classical control field with Rabi frequency $\Omega$ couples $\ket{s}$ to $\ket{e}$. A AFC is prepared on the inhomogeneously broadened $\ket{g} \rightarrow \ket{e}$ transition using techniques such as piecewise adiabatic passage or optical frequency comb excitation, with peak width $\gamma$, spacing $\Delta$, and total width $\Gamma$. The finesse is defined as $F = \Delta / \gamma$.

The cavity field obeys the Heisenberg–Langevin equation:
\begin{equation}
\partial_t \hat{E} = -\kappa \hat{E} + \sqrt{2\kappa} \hat{E}_{\mathrm{in}} + i g_0^2 \wp \int_{-\infty}^{\infty} d\delta\, n(\delta)\, \hat{\sigma}_{ge}(t; \delta),
\end{equation}
where $\kappa$ is the cavity decay rate, $g_0$ the atom–cavity coupling, $\wp$ the dipole matrix element, and $n(\delta)$ the atomic spectral distribution.

The atomic polarization evolves as:
\begin{equation}
\partial_t \hat{\sigma}_{ge} = -(i\delta + \gamma_p) \hat{\sigma}_{ge} + i \wp \hat{E},
\end{equation}
with $\gamma_p$ the homogeneous linewidth. The input–output relation is:
\begin{equation}
\hat{E}_{\mathrm{out}} - \hat{E}_{\mathrm{in}} = \sqrt{2\kappa} \hat{E}
\end{equation}

Defining collective spin operators for alkali and noble-gas ensembles:
\begin{align}
\hat{S}(\mathbf{r}, t) &= \frac{\hat{\sigma}_{gs}(\mathbf{r}, t)}{\sqrt{p_a n_a}}, \\
\hat{K}(\mathbf{r}, t) &= \frac{\hat{\sigma}_{\Downarrow\Uparrow}(\mathbf{r}, t)}{\sqrt{p_b n_b}},
\end{align}
the coupled evolution is:
\begin{subequations}\label{eq:PSK-system}
\begin{align}
\partial_t \hat{P} &= -(\gamma_p + i\bar{\delta})\hat{P} + i \Omega \hat{S} + i G \hat{E} + \hat{f}_P, \label{eq:Pdot} \\
\partial_t \hat{S} &= -(\gamma_s + i\delta_s - D_a \nabla^2)\hat{S} + i \Omega^* \hat{P} - i J \hat{K} + \hat{f}_S, \label{eq:Sdot} \\
\partial_t \hat{K} &= -(\gamma_k + i\delta_k - D_b \nabla^2)\hat{K} - i J \hat{S} + \hat{f}_K \label{eq:Kdot}
\end{align}
\end{subequations}

Initially, with $\Omega = 0$, the quantum light is absorbed by the alkali ensemble. Perfect impedance matching ($\kappa = Z$) ensures complete absorption:
\begin{equation}
\hat{E}_{\mathrm{out}} = \frac{\kappa - Z}{\kappa + Z} \hat{E}_{\mathrm{in}}, \quad Z = \frac{N_a g_0^2 \wp^2}{\Gamma}
\end{equation}

\begin{figure*}[t]           
  \centering
  \includegraphics[width=\textwidth]{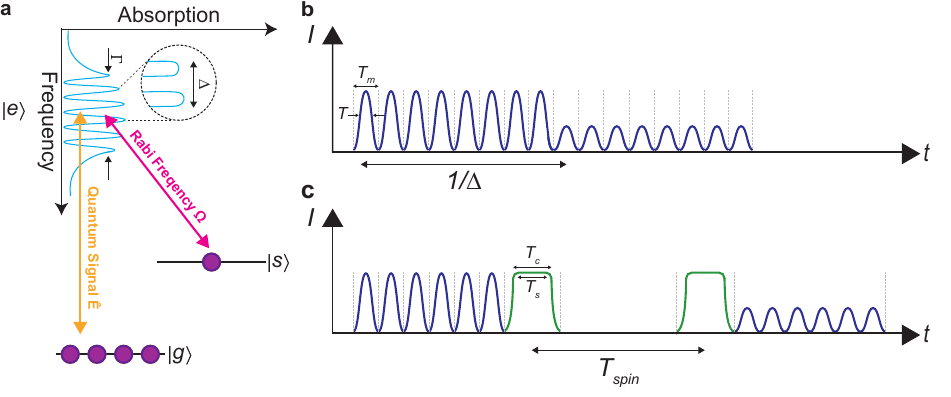}
\caption{\textbf{Atomic-frequency-comb (AFC) memory with fixed-delay and spin-wave readout.}
\textbf{(a)} In an inhomogeneously broadened transition, frequency-selective pumping carves a periodic absorption comb of spacing \(\Delta\) and total bandwidth \(\Gamma\). A weak quantum signal \(\hat{E}\) resonant with the comb is absorbed and re-emitted as an echo after the fixed rephasing time \(1/\Delta\) (two-level AFC). In the spin-wave variant, a control \(\pi\)-pulse coherently and reversibly transfers the excitation to a second ground state \(|s\rangle\), enabling on-demand readout and long storage times. 
\textbf{(b)} Time sequence for the fixed-delay AFC: the input is a train of temporal modes of duration \(T_m\), each mode containing pulses with intensity FWHM \(T\); the echo occurs at \(t=1/\Delta\) (or as a train for multiple inputs). 
\textbf{(c)} Time sequence for the AFC spin-wave memory: a control pulse of duration (bin size) \(T_c\) interrupts the rephasing and stores the excitation as a spin wave for a programmable time \(T_{\mathrm{spin}}\). Because the control occupies part of the AFC window, the allowable input-mode window is reduced to \(1/\Delta - T_c\); a second control pulse reconverts the spin wave, yielding an on-demand echo at \(t=1/\Delta+T_{\mathrm{spin}}\). Adapted from \cite{Ortu2022QST_AFCmultimode}.}

  \label{fig:overview}
  \vspace{-2mm}                  
\end{figure*}

The storage protocol proceeds in three main steps:

\begin{enumerate}
    \item \textbf{Optical-to-alkali spin transfer.}  
    Transfer the optical coherence to the alkali spin-wave $\widehat{S}$ using a chirped adiabatic pulse:
    \begin{equation}
        \langle \widehat{S}^\dagger \widehat{S} \rangle_{T} 
        = \left[ 1 - \exp\left( -\frac{\pi T \, \Omega^2}{\Gamma} \right) \right] 
        \langle \widehat{P}^\dagger \widehat{P} \rangle_{0}
    \end{equation}

    \item \textbf{Alkali-to-noble gas spin transfer.}  
    Transfer $\widehat{S}$ to the noble-gas spin $\widehat{K}$ via spin-exchange collisions:
    \begin{equation}
        \langle \widehat{K}^\dagger \widehat{K} \rangle_{T^{\prime}} 
        = \exp\left( -\frac{\pi \gamma_s}{2J} \right) 
        \langle \widehat{S}^\dagger \widehat{S} \rangle_{T}
    \end{equation}

    \item \textbf{Long-term storage.}  
    Store the excitation in $\widehat{K}$ for a duration limited by $1/\gamma_k$, which can potentially reach hundreds of hours.
\end{enumerate}

Reversing the sequence after a delay leads to re-emission after $2\pi/\Delta$:
\begin{equation}
\widehat{E}_{\mathrm{out}}(t) = -\sqrt{\eta_m} \, \widehat{E}_{\mathrm{in}}\left(t - 2T^{\prime} - 2T - \frac{2\pi}{\Delta}\right),
\end{equation}
with efficiency:
\begin{equation}
\eta_m = \left[ 1 - \exp\left( -\frac{\pi^2 T \Omega^2}{\Gamma} \right) \right]^2 \exp\left( -\frac{\pi \gamma_s}{J} \right) \mathrm{sinc}^2\left( \frac{\pi}{F} \right)
\end{equation}

\subsection{Satellite-Specific Modifications}
Deploying this memory in orbit requires proper consideration of the many spaceborne effects that would normally be absent in terrestrial deployments. First, the relative motion of a LEO satellite (with an orbital speed of several km/s) induces significant Doppler shifts in the optical frequencies, which are on the order of GHz for visible/near-IR wavelengths \cite{Dassie2023Doppler}. Without proper treatment, these Doppler offsets would detune the light from the alkali transitions and the cavity resonance. In practice, this is handled by real-time frequency tuning of the transmitter/receiver lasers based on the satellite’s ephemeris. Second, the free-space link coupling efficiency will be limited by diffraction and pointing errors. Even with diffraction-limited optics, a modest transmit aperture (e.g. 10–20 cm) yields a beam divergence on the order of \(10.0~\mu\text{rad}\), so the spot size at the ground can be several meters across for typical LEO altitudes \cite{Gundogan2021npjQI}. Coupling this free-space mode into the narrow mode of an optical cavity (or a single-mode fiber feeding the cavity) is lossy. Current designs of quantum transmitters estimate that only ~50–60\% of the collected light can be coupled into a single-mode channel, even under ideal alignment \cite{Ahmadi2024QUICK3}. Any pointing and tracking jitter exacerbates this loss. For example, a $\sim$\(2.0~\mu\text{rad}\) pointing error in a 20 cm downlink telescope can introduce an additional 4 dB loss in received power. The Micius quantum satellite demonstrated \(.60~\mu\text{rad}\) pointing accuracy using two-axis gimbaled telescopes and piezoelectric fast steering mirrors, which thus serve as the benchmark for our implementation \cite{Bedington2017Progress}. Specifically, our system would employ a similar multi-stage acquisition and tracking approach: open-loop slewing to point the telescope within a fraction of a degree, followed by a beacon-based fine pointing loop and fast electro-optic beam steering to correct residual errors \cite{Bedington2017Progress}. Lastly, atmospheric downlink effects must be considered. A downlink (satellite-to-ground) is generally preferred for quantum communication because the beam travels most of its path in the vacuum of free-space, with the dense atmosphere only being encountered at the end of the path. This avoids the dreaded “shower-curtain” effect of uplink channels, where turbulence near the transmitter greatly enlarges beam divergence early in the path \cite{Gundogan2021npjQI}. Despite this, downlink channels still experience high levels of atmospheric turbulence, which causes scintillation and wavefront distortions. Adaptive optics or feed-forward schemes may be used to partially correct the wavefront, but for a satellite transmitter with a small aperture, higher-order turbulence corrections yield only marginal gains \cite{Bedington2017Progress}.

\subsubsection{Microgravity Effects on Diffusion}
The space-based cavity will operate in microgravity, which alters the gas dynamics within the memory cell. On Earth, any thermal gradients or concentration gradients in the vapor cell can induce convective currents that mix the gas. In microgravity, this buoyant convection is suppressed, and transport becomes purely diffusive \cite{Beysens2022TransportMicrogravity}. The diffusion coefficients $D_a$ (alkali vapor) and $D_b$ (noble gas) themselves are functions of temperature and buffer gas pressure, and they are not fundamentally changed by gravity; however, the lack of convection means that these diffusion processes are the only means of distributing heat and mass in the cell. This has two main implications. First, without convective mixing, it may take longer for the alkali and noble-gas spin polarizations to reach a uniform steady state across the cell as spatial inhomogeneities must even out by diffusion alone. Second, any local heating (for instance, from the control laser or alkali pump light) will not be rapidly dissipated by fluid motion. Thus, thermal gradients could persist longer and could create spatial variations in $D_a$, $D_b$ or in spin relaxation rates. The cell design must, therefore, minimize thermal hotspots, which could be achieved by using thermal straps, radiative cooling, and a small magnetic stirring field, all of which promote mixing \cite{Niederhaus2008IVMixing,Selvadurai2022PassiveThermal}. Overall, microgravity offers a more stable environment, which is beneficial for maintaining atomic coherence; however, it places the burden of careful thermal management to ensure that diffusion is sufficient to homogenize the spin ensemble behavior throughout the duration of the storage time.

\subsection{Link Budget and Synchronization}
For a satellite-based deployment, the quantum memory must interface with the free-space optical link while synchronizing with the satellite’s visibility windows. The available contact time with a given ground station is limited to a few minutes per orbital pass (for LEO altitudes), so the protocol should efficiently multiplex and store as many photons as possible during each pass assuming line-of-sight is possible. The AFC-based multimode storage is well-suited for this: as illustrated in Fig.~\ref{fig:overview}, the memory can absorb a train of temporal modes within the fixed cavity delay $1/\Delta$. In practice, one would operate the satellite as it comes into line-of-sight with a ground transmitter by capturing successive entangled photon pulses into different temporal modes of the AFC. Immediately after absorption, these excitations are transferred to the long-lived nuclear spin ($\hat{K}$) via the two-step process described earlier. This buffering in the noble-gas spin allows the satellite to fly out of view while the quantum states are safely stored in spin coherence. Once the satellite reaches another ground station (or comes around for a second pass over the same station, if coherence time allows), a retrieval sequence (spin-wave transfer back to alkali, then optical read-out) is initiated to re-emit the photons towards the receiver. Using realistic parameters – for example, a 15 cm diameter transmit telescope on the satellite and a 50 cm receive telescope on the ground, with \(10.0~\mu\text{rad}\) beam divergence – one can expect on the order of $10^{-4}$ end-to-end efficiency (around 40 dB loss) for a high-elevation pass \cite{Bedington2017Progress, Zhang2018LargeScaleQKD}. This assumes clear weather and high pointing accuracy. Under such conditions, if the source on the ground sends, say, $10^6$ photons per second to the satellite, approximately 100 photons per second would enter the memory. Thanks to the multimodal capacity of the AFC (which can be in the range of $N \sim 10^3$ modes, the satellite could accumulate a large batch of entangled photons in each overpass. \cite{Gundogan2021npjQI}. The memory write/read efficiency and spin transfer efficiency then determine how many of those stored qubits can be retrieved with high fidelity. From this example, it is clear that high internal efficiency is essential for such a protocol to achieve high rates.

Therefore, our design targets near-unity as our goal for internal efficiency. To achieve this, we use impedance-matched absorption and optimal control pulses, as described by the theoretical efficiency $\eta_m$ derived above. By synchronizing the storage protocol with the satellite’s pass through the use of GPS timing and closed-loop feedback from the ground for the control pulses, we ensure that the memory is ready to receive photons at the start of the pass and that all stored modes can be read out on demand at the appropriate time. This synchronization extends to Doppler shift management: the ground station’s transmitter can pre-compensate the frequency based on the known satellite velocity profile so that the incoming signal is always at the memory’s resonance \cite{Dassie2023Doppler}. In addition, both ground and satellite will share synchronized clocks to align the write/read timing within the photon coherence time \cite{Bedington2017Progress}. Overall, these considerations ensure that the satellite memory can effectively integrate into a quantum network, capturing entanglement on the fly and releasing it when and where needed to perform entanglement swapping or to deliver a secret key.

\subsection{Space Payload Engineering Constraints}
Implementing the above capabilities on a satellite needs to be achievable, despite the payload constraints of a LEO satellite. We draw on proven strategies from ultra-cold atom experiments in space—notably, NASA’s Cold Atom Lab (CAL) aboard the International Space Station—to guide our design. CAL had to maintain an ultrahigh vacuum and stable conditions for Bose–Einstein condensates in orbit, which closely parallels the needs of our spaceborne quantum memory. In CAL’s design, a compact dual-chamber glass cell was used, with one science cell kept below $10^{-10}$ Torr by a combination of a miniaturized ion pump and non-evaporable getter (NEG) pumps \cite{Elliott2018CAL}. We similarly require a long-lived cell for the alkali–noble gas mixture; although our memory cell is sealed and does not need active pumping during operation, it must be processed to high vacuum standards to remove impurities and avoid excess alkali vapor pressure at operating temperature. A small getter pump can be included to absorb any outgassing over the mission lifetime. Surrounding the cell and cavity, magnetic shielding is essential. CAL employed a dual-layer metal magnetic shield to attenuate the strongly fluctuating magnetic fields on the ISS \cite{Elliott2018CAL}. Our satellite will encounter Earth’s magnetic field, as well as the spacecraft-induced fields. We believe a similar dual-layer shield can reduce ambient field exposure, which will ensure the spin precession frequencies ($\delta_s$, $\delta_k$) remain stable. Additional coil systems inside the shield can provide fine bias field control and enable rapid re-zeroing of fields if, for example, a magnetic disturbance from the satellite occurs. Vibration isolation and mechanical stability are also important factors to consider. Launch and deployment subject the payload to high vibration and shock. Even in orbit, reaction wheel jitter or thermal flexing could misalign the cavity. In CAL, all the delicate optics (mirrors, fiber collimators, cameras) were mounted on a solid aluminum support structure that was attached to the physics package \cite{Elliott2018CAL}. This ensured that any mechanical drift affecting all components was addressed and that the relative alignments were kept fixed. We adopt the same approach in our design: the cavity mirrors, coupling lenses, and detector optics will be integrated onto a single optical bench bonded to the vacuum package. Passive vibration damping, such as gel mounts or commercial shock absorbers, will protect the system during launch, and once in microgravity, the lack of seismic noise greatly benefits cavity stability. For the optical cavity specifically, we will use low-expansion materials and possibly small piezoelectric actuators to lock its length to the atomic resonance, counteracting thermal or mechanical drift.

Radiation in orbit poses another challenge. High-energy cosmic rays and solar particles can induce radiation damage or create errors such as bit flips in control electronics, or even spin flips in the atoms due to ionization events \cite{Binder1975SatelliteAnomalies}. CAL did not incorporate dedicated radiation shielding beyond the ISS hull; however, for our design, we must consider it. Specifically, our design uses radiation-hardened electronics (FPGA and microcontrollers rated for space) and places sensitive components (detectors, avalanche photodiodes, etc.) in light-weight shielding enclosures to block a significant portion of ionizing radiation \cite{Daneshvar2021MultilayerShield}. The alkali/noble gas atoms themselves are somewhat shielded by the vacuum cell walls and surrounding optics, but prolonged exposure to cosmic rays could lead to glass cell charging or diffusion of impurities. We will mitigate this by using a conductive coating on the cell to dissipate charges \cite{conductiveCoating}.

Thermal control is another crucial aspect that we must consider in our design. A satellite experiences rapid temperature swings as it moves in and out of sunlight. The hybrid memory, especially the vapor cell and cavity, must be kept within a narrow temperature range for optimal operation (typically, alkali vapor cells are heated to $\sim40$–$80,^\circ$C for sufficient vapor density, and the cavity must remain stable to much better than that to avoid drift in resonance). We implement a combination of passive and active thermal regulation. Externally, the payload is blanketed in multilayer insulation (MLI) to reduce radiative exchange with the environment, and heat is conducted to dedicated radiator panels that disperse waste heat to space. Internally, key components are thermally linked via high-conductivity straps (as in CAL’s water-cooled thermal straps for its coil assembly) to even out temperature gradients \cite{Elliott2018CAL}. Small resistance heaters with closed-loop PID control are attached near the cell and cavity to fine-tune their temperatures. On the CAL ISS experiment, fluid loops were used for cooling; however, but for our satellite, we prefer heat pipes or pumped liquid loops only if absolutely necessary due to mass and power overhead \cite{S3VI2025ThermalControl}. Instead, our design relies mainly on passive thermal stability, using the thermal mass of the optical bench and a low conductivity mounting to the satellite chassis to decouple fast fluctuations. All told, these engineering measures build upon the CAL experiment and other space quantum hardware demonstrations to ensure the quantum memory operates as intended in space.

\subsection{Implementation}
Following the original implementation from \cite{ogCRYOpaper,Katz2022QuantumInterface}, we consider a mixed $^3$He–K ensemble confined in an uncoated spherical glass cell of radius $R=1$ cm at $215^{\circ}$C. We operate in a high pressure configuration with working densities $n_a=5.0\times10^{14}\,\mathrm{cm}^{-3}$ for potassium and $n_b=6.0\times10^{19}\,\mathrm{cm}^{-3}$ for $^3$He \cite{Shaham2022StrongCoupling}. The coherent spin exchange coupling used in the simulations is set to $J=2.00\times10^{-5}$ \cite{foundationalPaper}. A nitrogen buffer gas at $30$ Torr is included for collisional quenching, which broadens the K D$_1$ transition to a linewidth of $\Gamma\approx27\,\mathrm{GHz}$. The AFC tooth width $\gamma$ is limited by the optical coherence lifetime and is bounded by twice the natural linewidth $\gamma_p=5.96\times2\pi\,\mathrm{MHz}$. Targeting a finesse $F=8$ gives a tooth spacing $\Delta=96\,\mathrm{MHz}$ \cite{foundationalPaper}. 

Assuming the AFC bandwidth matches the pressure broadened line, the multimode capacity is
\[
M=\frac{2\Gamma}{5\Delta}=\frac{2\times27~\mathrm{GHz}}{5\times96~\mathrm{MHz}}=112.
\]

For our quantum memory simulation, we use coupled spin–diffusion dynamics with Dirichlet boundary conditions for the alkali component and Neumann for the noble gas. When operating the satellite abroad, we expect a slightly higher steady state gas temperature due to the requirements of a stable thermal management system aboard. Due to advancements in current thermal management systems, we see this being minimal, as seen in \cite{Liu2018NatComm}. In addition to this, the pressure differential across the optical cavity seals and the outside environment increases in orbit relative to the ground. In turn, we expect a modest reduction in the fill pressure of the cavity. Because the diffusion of a dilute gas scales approximately as $D \propto T^{\frac{3}{2}}/P$ according to Chapman Enskog theory \cite{Gensterblum2015GasTransportShale}, we increase the diffusion constants by 10\% from the original paper, which gives  $D_a=1.02\times10^{-8}\,\mathrm{m^2/s}$ and $D_b=2.05\times10^{-8}\,\mathrm{m^2/s}$ respectively. For small detunings, we set $\delta_s=0$ and $\delta_k=1.11\times10^{-3}$ \cite{foundationalPaper}. For longitudinal relaxation rates, we set $\gamma_k=0$  and $\gamma_s=3.1\times10^{-7}\,\mathrm{s^{-1}}$ \cite{foundationalPaper}. All values were derived from the original study, with some updated to reflect the fact that this memory will be operating in space. 

Building a satellite-borne alkali–noble gas quantum memory will require the integration of the above considerations into a satellite-ready system where miniaturization and integration will be key. Fortunately, the core of the memory—a vapor cell and optical cavity—can be quite compact. For example, the entire dual-cell vacuum chamber used in CAL had dimensions on the order of a few centimeters, and compact vapor cells with integrated nano-coating technology have been demonstrated for atomic clocks and magnetometers \cite{Elliott2018CAL,Wang2022MicroFabricatedVaporCells}. Utilizing recent developments in integrated photonics and micro-optics, the cavity could be formed by tiny high-reflectivity mirrors directly bonded to the cell or on a ceramic spacer, and semiconductor diode lasers (in the 780–800 nm range) can be made flight-qualified and small. The control and readout lasers might be fiber-coupled distributed Bragg reflector (DBR) lasers, locked via spectroscopy to an onboard reference cell. All laser beams can be delivered through polarization-maintaining fibers to the memory module, similar to the fiber-fed design of CAL’s optical system \cite{Elliott2018CAL}. The entire science module, which contains the memory cell, cavity, minor optical elements, and shielding – would be mounted on a small optical bench (~30 cm) and could fit inside a microsatellite (50–100 kg class) or possibly even a 12U CubeSat with careful design. The electronics (RF coils for driving spin transitions, laser drivers, single-photon detectors, and a processor for control logic) would occupy another section of the satellite bus, which is designed with redundancy and fail-safes typical of space missions.

\section{System Model}
\label{sec:system-model}
This section defines our space-to-ground (S–G) link model and how we quantified the success of our memory proposal. We factor the per-trial success probability into source, memory, detector, atmospheric, and diffraction/pointing terms, each with explicit elevation/slant-range dependence. We then compute SKR under BB84 with one-way post-processing and compare two geometries: (i) dual downlink at fixed low elevation, and (ii) buffered overhead with zenith links subject to \(T_{\mathrm{mem}}\!\ge\!t_{\mathrm{buffer}}\).

\subsection{Link Budget}
\label{subsec:link-budget-sg}

We model each S-G elementary link as a product of statistically independent transmission factors. The single-shot success probability is
\begin{equation}
\eta_{\mathrm{S\text{--}G}}
=
\eta_{\mathrm{src}}\;
\eta_{\mathrm{mem}}\;
\eta_{\mathrm{det}}\;
\eta_{\mathrm{atm}}(\theta)\;
\eta_{\mathrm{dif}}(l),
\end{equation}
where \(\eta_{\mathrm{src}}\) is the entangled-photon source efficiency, \(\eta_{\mathrm{mem}}\) the memory write/read efficiency on the satellite, \(\eta_{\mathrm{det}}\) the (ground) detector efficiency, \(\eta_{\mathrm{atm}}(\theta)\) the single-pass atmospheric transmission at elevation \(\theta\), and \(\eta_{\mathrm{dif}}(l)\) the diffraction\,+\,pointing coupling over slant range \(l\).
For the configurations studied here we use
\begin{equation}
\eta_{\mathrm{src}}=0.20,\qquad
\eta_{\mathrm{mem}}=0.74,\qquad
\eta_{\mathrm{det}}=0.70.
\end{equation}

\paragraph{Atmospheric transmission.}
Single-pass attenuation along the slant path at elevation \(\theta\) follows a simple air-mass model:
\begin{equation}
\eta_{\mathrm{atm}}(\theta)
=
\bigl(\eta_{\mathrm{atm}}^{z}\bigr)^{1/\sin\theta},
\end{equation}
where \(\eta_{\mathrm{atm}}^{z}\) is the zenith transmission (set by wavelength and site conditions).

\paragraph{Diffraction and pointing.}
Collected power at the ground telescope is expressed as
\begin{equation}
\eta_{\mathrm{dif}}(l)=\frac{P(l)}{P_0}.
\end{equation}
A diffraction-limited Gaussian beam of waist \(w_0\) obeys
\begin{equation}
\theta_d=\frac{\lambda}{\pi w_0},\qquad
w(z)=w_0\sqrt{1+\Bigl(\frac{\theta_d z}{w_0}\Bigr)^2},
\end{equation}
\begin{equation}
I(r,z)=I_0\Bigl(\frac{w_0}{w(z)}\Bigr)^2\exp\!\Bigl[-\frac{2r^2}{w(z)^2}\Bigr].
\end{equation}
Residual pointing jitter is modeled by a 2D Gaussian kernel with rms \(\sigma_p\):
\begin{equation}
g_p(\mathbf{r})=\frac{1}{2\pi\sigma_p^2}\exp\!\Bigl(-\frac{\|\mathbf{r}\|^2}{2\sigma_p^2}\Bigr),
\end{equation}
and the received intensity is the convolution \(\tilde I = I * g_p\). Power coupled into a circular pupil of radius \(R_{\mathrm{rec}}\) at range \(l\) is
\begin{equation}
P(l)=\int_{0}^{R_{\mathrm{rec}}}\!\rho\,d\rho\int_{0}^{2\pi}\!d\varphi\;\tilde I(\rho,l).
\end{equation}

\paragraph{Geometry.}
The elevation \(\theta\) and slant range \(l\) are determined from the ground-track separation \(L_g\) and orbital altitude \(h\):
\begin{equation}
l^2
=
R_E^2 + (R_E+h)^2 - 2R_E(R_E+h)\cos\!\Bigl(\frac{L_g}{R_E}\Bigr),
\end{equation}
\begin{equation}
\theta
=
\frac{\pi}{2}
- \frac{L_g}{R_E}
- \arcsin\!\Biggl[\frac{R_E}{l}\sin\!\Bigl(\frac{L_g}{R_E}\Bigr)\Biggr],
\end{equation}
with \(R_E=6{,}371~\mathrm{km}\). For a more detailed overview of the parameters used in this study, see Appendix \ref{app:parameters}.

\section{Results \& Analysis}
\label{sec:results}
We first visualize the memory dynamics, then quantify SKR gains, and finally compare the two different operational scenarios.

\begin{figure*}[!t]
  \centering
  \includegraphics[width=\textwidth]{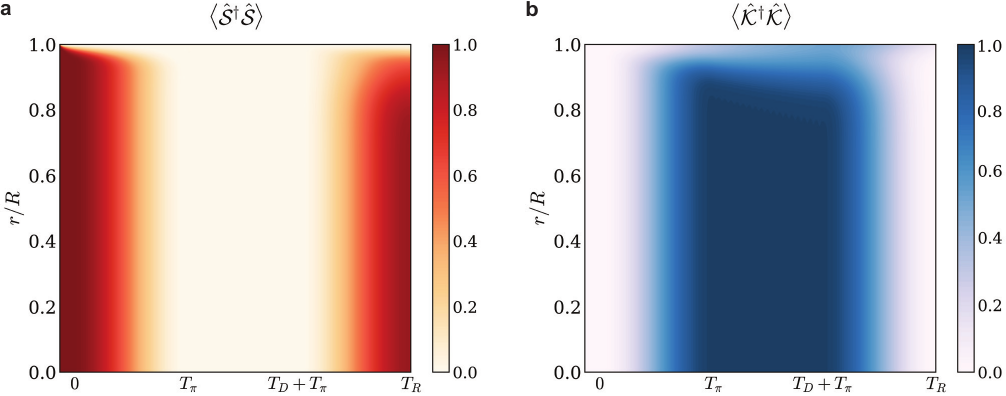}
  \caption{\textbf{Spatiotemporal kymographs of the collective spins.}
    \textbf{(a)} Alkali–spin kymograph: color encodes the normalized magnitude \(\langle \hat S^{\dagger}\hat S\rangle(r,t)\) on a radial slice of the cell where $r$ is the distance from the center and $R$ is the cell radius. The normalized radial slice of the cell runs from $r/R\in [0,1]$ where 0 is the center and 1 is the wall which is shown over time. The bright bands at the write and read instants \(T_{\pi}\) and \(T_{R}\) indicate loading and retrieval of the alkali spin wave; attenuation during the dark interval \(T_{D}\) reflects homogeneous decay \(\gamma_s\) and diffusion \(D_a\).
    \textbf{(b)} Noble-gas–spin kymograph: color encodes \(\langle \hat K^{\dagger}\hat K\rangle(r,t)\). Population builds during the alkali\(\rightarrow\)noble-gas transfer at \(T_{\pi}\) via spin-exchange \(J\), remains high throughout storage with mild edge losses from wall relaxation and diffusion \(D_b\), and then maps back at \(T_{R}\).
    Both panels are obtained by solving the coupled spin–diffusion equations (Eqs.~\eqref{eq:Pdot}--\eqref{eq:Kdot}) and plotting the azimuthally averaged \(|S(r,t)|^{2}\) and \(|K(r,t)|^{2}\), each normalized to the initial optical excitation.}
  \label{fig:overview}
\end{figure*}

\subsection{Downlink Link Establishment Probability}
\label{subsec:results-downlink}

\begin{figure}[H]
  \centering
  \includegraphics[width=\columnwidth]{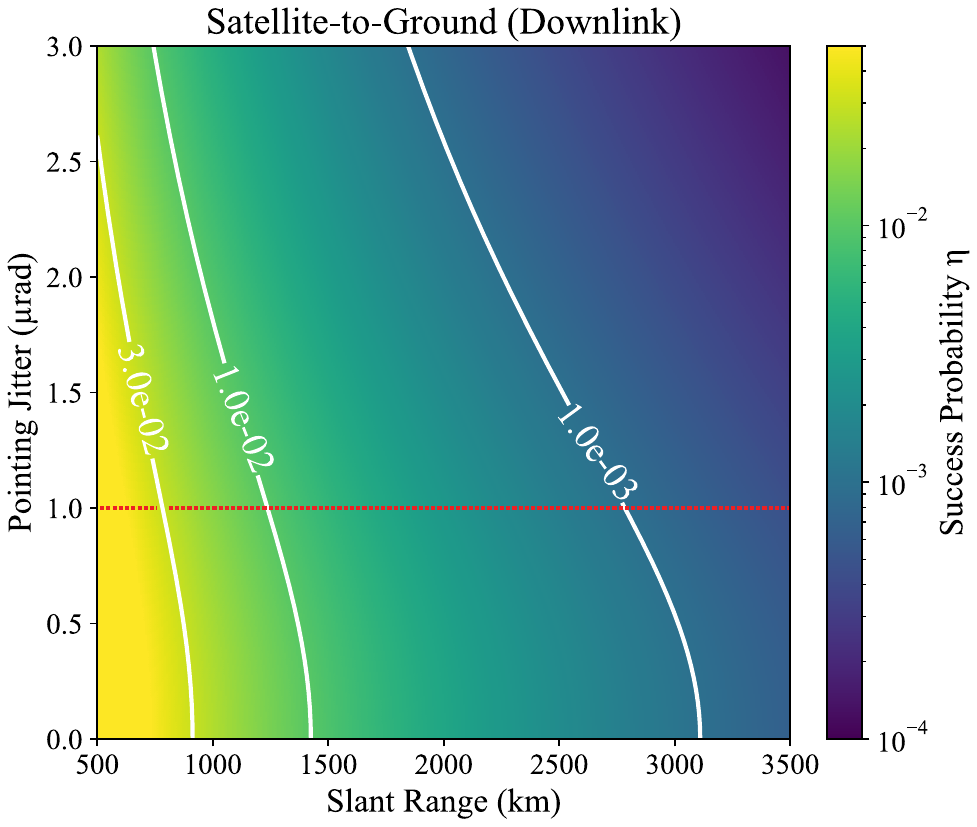}
  \caption{\textbf{Satellite–to–ground downlink success probability.}
Color indicates the single-photon success probability \(\eta\) as a function of slant range (horizontal axis) and rms pointing jitter (vertical axis); white curves are iso-\(\eta\) contours. Larger range and/or jitter reduce \(\eta\) due to diffraction and pointing loss. The red dashed line marks the operating assumption for pointing-jitter used in this work (rms of \(1.0~\mu\text{rad}\)).}
  \label{fig:downlink_prob}
\end{figure}

Figure~\ref{fig:downlink_prob} maps the single-photon downlink success probability \(\eta\) as a function of slant range and rms pointing jitter. Two features dominate: (i) diffraction loss with increasing range and (ii) the strong penalty from microradian-level jitter, which broadens the far-field spot relative to the receive pupil. Operating near zenith (short range) substantially mitigates both effects.

\subsection{Memory Efficiency}
\label{subsec:mem-eff}

Utilizing the code from \cite{foundationalPaper}, we compute the expectation values of the collective spins for the alkali and noble-gas ensembles over the full protocol by numerically integrating the coupled spin equations (Eqs.~\eqref{eq:Pdot}--\eqref{eq:Kdot}). Spatial discretization uses the method of lines, and time integration employs a time-adaptive Runge--Kutta scheme. The boundary conditions for these equations follow the physical surface interactions of the two species. At the glass–vapor interface, the alkali polarization is set to zero (Dirichlet): 
\[
S(r=R,t)=0.
\]
This destructive boundary condition properly accounts for the fact that alkali atoms undergo strong depolarizing collisions with uncoated cell walls, which scramble the electronic spin and prevent coherent spin-wave amplitude from surviving at the boundary \cite{Chang2024WallCollisionEffect, Wu2021WallInteractionsSpinPolarizedAtoms}. As a result, alkali coherence decays toward the wall and is effectively quenched at $r=R$. Thus, from this, we see that the spatial spin-wave mode will be naturally localized toward the central region of the cell.

In contrast, the noble-gas nuclear spin exhibits extremely weak depolarization in collisions with clean glass walls, so that wall limited relaxation times approach hundreds of hours even in small cells \cite{Gentile2017OpticallyPolarizedHe3}. To capture this, we impose a von Neumann boundary condition on the noble-gas spin, which therefore meets the following condition: 
\[
\partial_r K(r,t)\big|_{r=R}=0.
\]
This corresponds to a non-destructive, reflective boundary where no polarization current flows through the wall. Physically, this enforces that the noble-gas polarization remains smooth across the boundary and that the wall does not act as a relaxation sink, which allows coherence to remain stable for hundreds of hours, as we observe in experiments. 

\subsection{Instantaneous SKR (Space-to-Ground)}
\label{subsec:inst-skr-sg}

We quantify the instantaneous secret key rate (SKR) for a S-G entanglement distribution link using the standard asymptotic lower bound for BB84 with one–way post–processing. Per channel use, the secret key fraction is
\begin{equation}
R \;=\; \frac{Y}{2}\,\bigl[\,1 - h(e_{X}) - f\,h(e_{Z})\,\bigr],
\end{equation}
where \(Y\) is the success probability per channel use that the protocol is heralded (e.g., both parties register a valid detection and the Bell-state measurement succeeds), \(e_{X}\) and \(e_{Z}\) are the quantum bit error rates (QBERs) in the \(X\) and \(Z\) bases, and \(f\!\ge\!1\) captures error-correction inefficiency. This expression follows Refs.~\cite{Panayi2014NJP,Luong2016APB,Gundogan2021npjQI}. The binary entropy function is
\begin{equation}
h(e) \;=\; -\,e\log_{2}e \;-\; (1-e)\log_{2}(1-e).
\end{equation}
Details of how \(Y\) and the individual error terms are computed for the link model are provided in Refs.~\cite{Panayi2014NJP,Ma2007EntangledSourcesQKD,Trenyi2020MQMBeatingDirect}.

Given a channel–use rate (e.g., pair-generation or pulse) \(\nu\) [s\(^{-1}\)], the instantaneous SKR is
\begin{equation}
R_{\mathrm{inst}}(t) \;=\; \nu(t)\; R(t),
\end{equation}
which may vary over a pass through its time dependence in the elevation angle(s), losses, and background.

\paragraph{Yield decomposition.}
For the S–G geometry, we write
\begin{equation}
Y(t) \;=\; P_{\mathrm{herald}}\;
\eta_{A}(t)\,\eta_{B}(t),
\end{equation}
where \(P_{\mathrm{herald}}\) is the protocol-specific heralding probability (e.g., Bell-state measurement success), and \(\eta_{A}(t),\eta_{B}(t)\) are the end-to-end transmission efficiencies from the satellite to each optical ground station (OGS), including diffraction, atmosphere, pointing, receiver optics, detector efficiency, and (when used) memory efficiency. The elevation-angle dependence enters through the slant path \(L_{s}(\theta)\) and the corresponding atmospheric and geometric losses.

\paragraph{Scenarios considered.}
We evaluate two operational cases that differ only in pointing geometry and the use of buffering:

\begin{enumerate}
\item \textbf{Dual–downlink.} A single satellite distributes entangled photon pairs (EPS) simultaneously to two OGS, each tracked at a fixed low elevation, here \(\theta_{A}=\theta_{B}=20^{\circ}\). The longer slant ranges \(L_{s}(20^{\circ})\) reduce \(\eta_{A,B}\), so that
\begin{equation}
\begin{aligned}
Y_{\mathrm{dual}}(t)
&= P_{\mathrm{herald}}\,
   \eta_{A}\!\bigl(\theta_{A}\!=\!20^{\circ},t\bigr)\\
&\qquad\times
   \eta_{B}\!\bigl(\theta_{B}\!=\!20^{\circ},t\bigr),
\end{aligned}
\end{equation}

\begin{equation}
\begin{aligned}
R_{\mathrm{inst,dual}}(t)
&= \nu(t)\,\frac{Y_{\mathrm{dual}}(t)}{2}\,
   \bigl[\,1 - h(e_{X}) - f\,h(e_{Z})\,\bigr].
\end{aligned}
\end{equation}

\item \textbf{Buffered downlink.}
A single satellite performs two \emph{sequential} zenith downlinks. First, it transmits to OGS A at \(\theta=90^{\circ}\); at zenith the slant range equals the orbital height,
\begin{equation}
L_{s}(90^{\circ}) \;=\; h .
\end{equation}
The detected (or heralded) qubit is then stored until a subsequent zenith opportunity to OGS B, at which point a second \(\theta=90^{\circ}\) downlink completes the entangled pair. If \(t_{\mathrm{buffer}}\) is the storage interval between these two zenith opportunities, the memory must satisfy
\begin{equation}
T_{\mathrm{mem}} \;\ge\; t_{\mathrm{buffer}} .
\end{equation}

The effective yield (including memory efficiency \(\eta_{\mathrm{mem}}\)) is
\begin{equation}
\begin{aligned}
Y_{\mathrm{buff}}(t)
&= P_{\mathrm{herald}}\,\eta_{\mathrm{mem}}\,
   \eta_{A}\!\bigl(\theta\!=\!90^{\circ},t_{\text{first}}\bigr)\\
&\quad\times
   \eta_{B}\!\bigl(\theta\!=\!90^{\circ},t_{\text{second}}\bigr),
\end{aligned}
\end{equation}
and the corresponding instantaneous SKR is
\begin{equation}
\begin{aligned}
R_{\mathrm{inst,buff}}(t)
&= \nu(t)\,\frac{Y_{\mathrm{buff}}(t)}{2}\,
   \bigl[\,1 - h(e_{X}) - f\,h(e_{Z})\,\bigr].
\end{aligned}
\end{equation}

\end{enumerate}

\subsection{Memory Efficiency and Multimode Gain}
\label{subsec:results-memory}

The satellite-borne memory enters the end-to-end yield multiplicatively; for the baseline considered here we use
\begin{equation}
\eta_{\mathrm{mem}}=0.74
\end{equation}
This value is obtained from the analysis shown in Fig. ~\ref{fig:overview} and is obtained by solving the full PDE system for alkali–noble-gas spin exchange and extracting the ratio of retrieved to input quantum-state magnitudes. In addition, the memory supports \(N=112\) independent temporal modes. For a per-mode success \(P\ll 1\), the probability of at least one success within a cycle becomes \cite{ogCRYOpaper}
\begin{equation}
P_{N} \;=\; 1-(1-P)^{N} \;\approx\; N\,P \qquad (N=112),
\end{equation}
so the buffered downlink geometry benefits from an \(\approx 112\times\) multimode gain in the low-probability regime, compounding the high-elevation link advantage until clocking or readout limits are reached \cite{simon2007quantum,foundationalPaper}.

\subsection{Instantaneous SKR}
\label{subsec:results-skr}

Under the BB84 one-way post-processing bound (Sec.~\ref{subsec:inst-skr-sg}), the buffered downlink configuration achieves an instantaneous rate improvement of
\begin{equation}
\mathcal{G} \;\equiv\; \frac{R_{\mathrm{inst,buff}}}{R_{\mathrm{inst,dual}}} \;=\; 111 \times
\end{equation}
This gain is driven primarily by the reduced zenith slant path (\(500~\mathrm{km}\) versus \(1461.9~\mathrm{km}\)) and persists despite finite \(\eta_{\mathrm{mem}}=0.74\). The buffer interval required by the geometry is \(t_{\mathrm{buffer}}=463~\mathrm{s}\), so the operational constraint is \(T_{\mathrm{mem}}\!\ge\!463~\mathrm{s}\).

\begin{figure}[h]
  \centering
  \includegraphics[width=\columnwidth]{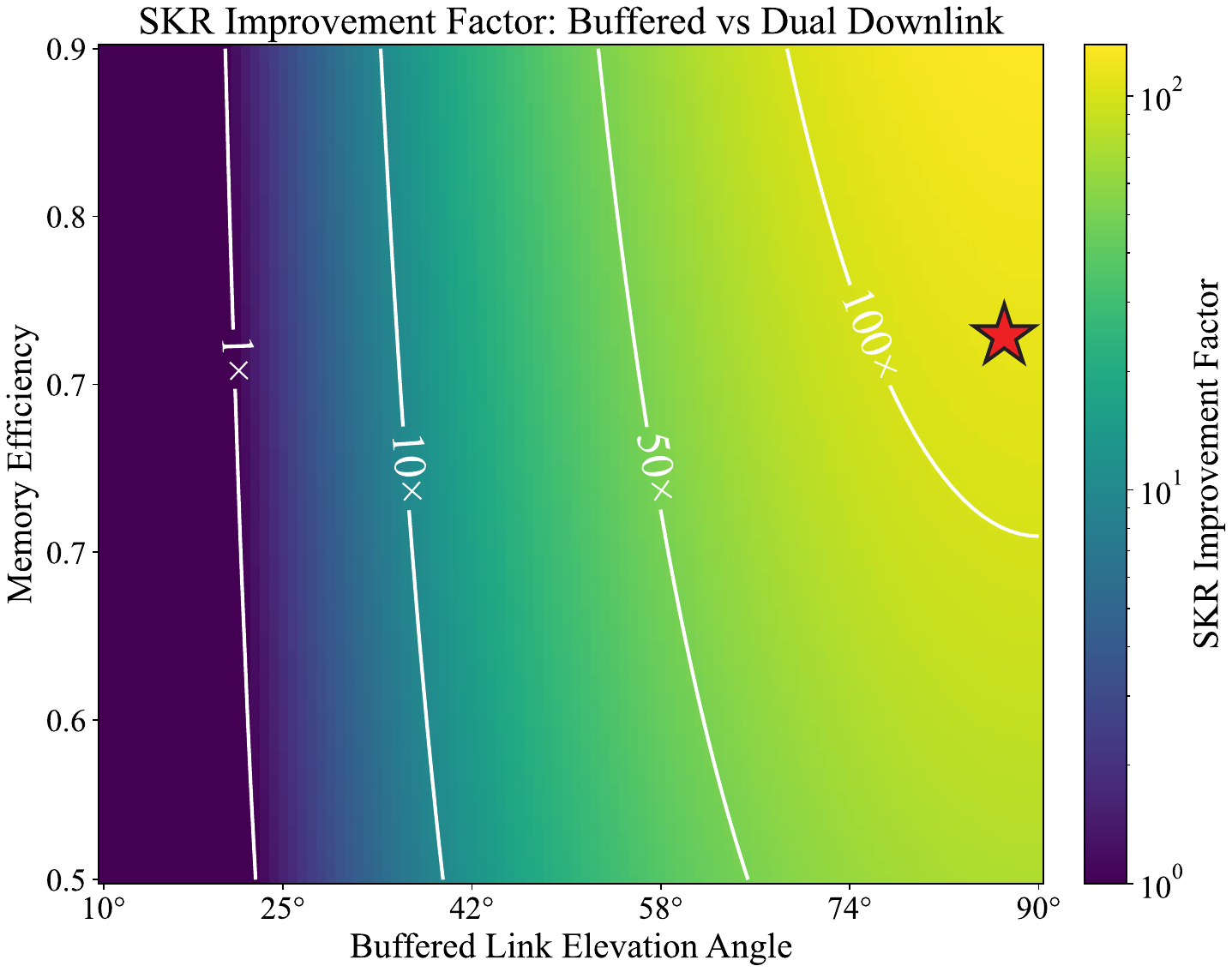}
  \caption{\textbf{SKR improvement for buffered downlink vs.\ dual downlink.}
  Heatmap of the instantaneous SKR gain \( \mathcal{G}\equiv R_{\mathrm{inst,buff}}/R_{\mathrm{inst,dual}} \) as a function of buffered-link elevation angle (x–axis) and memory efficiency \(\eta_{\mathrm{mem}}\) (y–axis). Colors (log scale) and iso-contours (1\(\times\), 10\(\times\), 50\(\times\), 100\(\times\)) show how operating nearer to zenith and improving \(\eta_{\mathrm{mem}}\) increase the advantage of buffering. The red star marks the baseline used in this work (\(\eta_{\mathrm{mem}}\approx0.74\), \(\theta\approx90^{\circ}\)), which yields a \(\gtrsim 10^{2}\!\times\) improvement.}
  \label{fig:skr_improve}
\end{figure}

\subsection{Scenario Comparison}
\label{subsec:scenario-comparison}

Table~\ref{tab:scenario-comparison-singlecol} summarizes the combined per-trial success probability and instantaneous SKR for the two downlink geometries at \(h=500~\mathrm{km}\) with an OGS separation of \(3267.9~\mathrm{km}\).

\begin{table}[t]
  \caption{Combined per-trial success probability \(\eta\) and instantaneous secret key rate (SKR) at \(h=500~\mathrm{km}\) and OGS separation \(3267.9~\mathrm{km}\).}
  \label{tab:scenario-comparison-singlecol}
  \squeezetable
  \begin{ruledtabular}
  \begin{tabular}{lcc}
    Scenario            & Combined \(\eta\)        & SKR (bits/s)        \\
    \hline
    Dual downlink       & \(4.23\times 10^{-5}\)   & \(8.12\times 10^{3}\) \\
    Buffered downlink   & \(4.70\times 10^{-3}\)   & \(9.03\times 10^{5}\) \\
    \hline
    \textbf{Improvement factor} & \(\bm{111\times}\) & \(\bm{111\times}\) \\  \end{tabular}
  \end{ruledtabular}
\end{table}

For these mission parameters, buffered downlink increases the combined success probability and instantaneous SKR by \(111\times\). The required storage interval between zenith contacts is \(t_{\mathrm{buffer}}=7.7~\mathrm{min}=463~\mathrm{s}\); feasibility therefore requires
\begin{equation}
T_{\mathrm{mem}} \;\ge\; 463~\mathrm{s}.
\end{equation}

\section{Discussion}
\label{sec:discussion}

\subsection{Buffered Overhead as an Alternative to Dual Downlink}
The results indicate that operating each downlink at zenith with intermediate storage constitutes a compelling alternative method for simultaneous dual downlink. For the baseline geometry (\(h=500~\mathrm{km}\), OGS separation \(3267.9~\mathrm{km}\)), the buffered-overhead approach shortens the slant range from \(1461.9\) to \(500.0~\mathrm{km}\) (\(2.9\times\)) and yields a \(111\times\) increase in both combined success probability and instantaneous SKR relative to the \(20^{\circ}/20^{\circ}\) case. This improvement arises primarily from reduced diffraction and pointing loss at zenith, which is consistent with the steep \(\eta\)-versus-range/jitter gradients observed in Fig.~\ref{fig:downlink_prob}. While the buffered scheme introduces a storage interval between zenith contacts, the required buffer time of \(t_{\mathrm{buffer}}=463~\mathrm{s}\) is compatible with the memory parameters considered here.

\subsection{Near-Term Feasibility of Satellite Quantum Memories}

Our calibrated memory efficiency \(\eta_{\mathrm{mem}}=0.74\) and support for \(N=112\) temporal modes suggest that meaningful SKR gains are achievable with near-term hardware. In the long-range, low-per-attempt success regime, multimode operation increases the probability of at least one success per cycle approximately linearly with \(N\) \cite{simon2007quantum,foundationalPaper}, which in turn enhances the elevation-angle advantage provided by the buffered geometry. The feasibility condition \(T_{\mathrm{mem}}\!\ge\!t_{\mathrm{buffer}}=463~\mathrm{s}\) sets a concrete requirement for storage lifetime. The storage decay over this interval is sufficiently small to justify treating \(\eta_{\mathrm{mem}}\) as constant in the SKR evaluation.

From a systems standpoint, the buffered-overhead approach is ideal because it trades stringent pointing at long slant range for a memory-limited duty cycle at short slant range. This reduces sensitivity to microradian-level jitter, which is the dominant penalty in Fig.~\ref{fig:downlink_prob}) and shifts engineering effort toward memory stability, control-pulse fidelity, and timing synchronization—all subsystems that can be validated on the ground and subsequently space-qualified. 

\subsection{Payload Integration and Scaling}
A practical advantage of this proposal is the physical compactness of the optical cavity. The memory modules considered here have \(\sim\!1~\mathrm{cm}\) dimensions which enables multiple independent memories per satellite without prohibitive mass/volume impact. Such replication supports:
\begin{enumerate}
\item \emph{Parallelism:} multiple memories can time-stagger or frequency-division multiplex downlinks to increase effective mode capacity beyond \(N=112\).
\item \emph{Redundancy:} cold spares mitigate radiation-induced performance drift and extend mission lifetime.
\item \emph{Functional partitioning:} specialized memories (e.g., longer-storage units dedicated to buffering vs.\ high-throughput units for bursty operation) allow adaptive scheduling across a pass.
\end{enumerate}
These architectural advantages further strengthen the buffered-overhead strategy and create a path to scaling SKR without relying solely on aperture growth or ultra-stable pointing.

\section{Conclusion \& Future Work}
\label{sec:conclusion}

Our results show that buffered zenith operations can deliver a \(\sim\!111\times\) increase in instantaneous SKR relative to simultaneous dual downlink at \(20^{\circ}\) elevation for the same protocol parameters, which is enabled by shorter zenith slant path (\(500~\mathrm{km}\) vs.\ \(1462~\mathrm{km}\)) and quantum memory parameters (\(\eta_{\mathrm{mem}}=0.74\), \(N=112\) temporal modes). The required storage lower bound for this buffer \(t_{\mathrm{buffer}}=463~\mathrm{s}\) is compatible with near-term memory performance, motivating buffered overhead as a primary mode for space–to–ground QKD demonstrations.

To build on this work, we outline several research directions that would be interesting to explore: 
\begin{enumerate}
\item \textbf{End-to-end power and mass budget.} Perform a full power budget and thermal analysis to size the spacecraft bus (solar array area, battery capacity, radiator sizing) required to operate all of the required hardware components. A key output of this study would be the estimation of how many memory modules can be hosted per satellite within the mass, volume, and power constraints.
\item \textbf{Expanded network simulations.} Extend from single-hop S–G links to multi-repeater constellations, quantifying secret-key throughput for a set experimental time window.
\item \textbf{Memory physics and coherence.} Refine device-level models to determine realistic coherence times (including radiation, temperature gradients, wall interactions, and control noise), and assess whether long-\(T_{\mathrm{mem}}\) operation can buffer gaps when line-of-sight to an OGS is temporarily unavailable due to cloud cover.
\end{enumerate}

Together, these efforts will help to better inform how the satellite architecture proposed in this study needs to be adapted to support a real-world implementation of this technology. 

\section{Data Availability}

The datasets and code supporting the findings of this study are publicly available in the GitHub repository at the following  \href{https://github.com/connor-a-casey/quantum-memory-satellite}{link}. This repository includes all necessary scripts, input data, and results to replicate and build upon the analyses presented in this paper. For questions or potential collaboration opportunities, please reach out to the corresponding author.

\bigskip

\section{Acknowledgments}
We would like to express our gratitude to Professor Don Towsley, as well as Prateek Mantri and Stav Halder, for their assistance and feedback as we drafted the manuscript and implemented the simulation model. 

\section{Contributions}
C.C. initiated and led the project, formulating the research question, designing the simulation methodology, implementing the complete codebase, and drafting most of the manuscript with input from all authors. A.W. assisted in modeling the network, proposed the idea of an instantaneous SKR, and provided extensive feedback on the simulation methodology. C.M. assisted with the literature review and the writing of the manuscript. E.R assisted in the overall flow of the manuscript, wrote a bulk of the background section, and provided feedback on the proposed satellite design. N.D. assisted in the literature review, as well as the writing of the manuscript. 

\section{Competing Interests}
The authors declare no competing interests.

\bibliography{citations}

@article{nature1650,
  author    = {S. H. Wei and B. Jing and X. Y. Zhang and J. W. Liu and K. Y. Liu and Y. Tian and X. J. Liu and J. H. Wu and J. Wu and W. Chen and X. M. Jin},
  title     = {Quantum storage of 1650 modes of single photons at telecom wavelength},
  journal   = {npj Quantum Information},
  year      = {2024},
  volume    = {10},
  number    = {},
  pages     = {19},
  doi       = {10.1038/s41534-024-00812-1},
  url       = {https://doi.org/10.1038/s41534-024-00812-1}
}

@article{DLCZ,
  author    = {L. M. Duan and M. Lukin and J. I. Cirac and P. Zoller},
  title     = {Long-distance quantum communication with atomic ensembles and linear optics},
  journal   = {Nature},
  volume    = {414},
  pages     = {413--418},
  year      = {2001},
  doi       = {10.1038/35106500},
  url       = {https://doi.org/10.1038/35106500}
}

@article{foundationalPaper,
  author = {Alexandre Barbosa and Hugo Terças and Emmanuel Zambrini Cruzeiro},
  title = {Multiplexed quantum repeaters with hot multimode alkali-noble gas memories},
  journal = {arXiv preprint arXiv:2402.17752},
  year = {2024},
  archivePrefix = {arXiv},
  eprint = {2402.17752},
}

@misc{gundogan2021quantum,
  author       = {Mustafa Gündoğan and Thomas Jennewein and Faezeh Kimiaee Asadi and Elisa Da Ros and Erhan Sağlamyürek and Daniel Oblak and Tobias Vogl and Daniel Rieländer and Jasminder Sidhu and Samuele Grandi and Luca Mazzarella and Julius Wallnöfer and Patrick Ledingham and Lindsay LeBlanc and Margherita Mazzera and Makan Mohageg and Janik Wolters and Alexander Ling and Mete Atatüre and Hugues de Riedmatten and Daniel Oi and Christoph Simon and Markus Krutzik},
  title        = {Topical White Paper: A Case for Quantum Memories in Space},
  year         = {2021},
  doi          = {10.48550/arXiv.2111.09595},
  url          = {https://doi.org/10.48550/arXiv.2111.09595}
}

@book{nc,
  author = {Michael A. Nielsen and Isaac L. Chuang},
  title = {Quantum Computation and Quantum Information: 10th Anniversary Edition},
  year = {2010},
  publisher = {Cambridge University Press},
  address = {Cambridge},
}

@article{martin2021quantum,
  author       = {Vivien Martin and Juan Carlos Garcia-Escartin and Christoforos N. Gagatsos and Enrico Diamanti and Philip Walther and Anthony Leverrier and Stefano Pirandola and et al.},
  title        = {Quantum technologies in the telecommunications industry},
  journal      = {EPJ Quantum Technology},
  volume       = {8},
  number       = {1},
  pages        = {19},
  year         = {2021},
  doi          = {10.1140/epjqt/s40507-021-00108-9},
  url          = {https://doi.org/10.1140/epjqt/s40507-021-00108-9}
}

@article{deforges2023satellite,
  author       = {L. de Forges de Parny and O. Alibart and J. Debaud and et al.},
  title        = {Satellite-based quantum information networks: use cases, architecture, and roadmap},
  journal      = {Communications Physics},
  volume       = {6},
  pages        = {12},
  year         = {2023},
  doi          = {10.1038/s42005-022-01123-7},
  url          = {https://doi.org/10.1038/s42005-022-01123-7}
}

@article{vinay2017practical,
  author       = {S. E. Vinay and P. Kok},
  title        = {Practical repeaters for ultralong-distance quantum communication},
  journal      = {Physical Review A},
  volume       = {95},
  pages        = {052336},
  year         = {2017},
  doi          = {10.1103/PhysRevA.95.052336},
  url          = {https://doi.org/10.1103/PhysRevA.95.052336}
}

@article{sit2017high,
  author       = {A. Sit and F. Bouchard and R. Fickler and J. Gagnon-Bischoff and H. Larocque and K. Heshami and D. Elser and C. Peuntinger and K. Günthner and B. Heim and C. Marquardt and G. Leuchs and R. W. Boyd and E. Karimi},
  title        = {High-dimensional intracity quantum cryptography with structured photons},
  journal      = {Optica},
  volume       = {4},
  number       = {9},
  pages        = {1006--1010},
  year         = {2017},
  doi          = {10.1364/OPTICA.4.001006},
  url          = {https://doi.org/10.1364/OPTICA.4.001006}
}

@article{yang2016efficient,
  author       = {S. J. Yang and X. J. Wang and X. H. Bao and et al.},
  title        = {An efficient quantum light--matter interface with sub-second lifetime},
  journal      = {Nature Photonics},
  volume       = {10},
  pages        = {381--384},
  year         = {2016},
  doi          = {10.1038/nphoton.2016.51},
  url          = {https://doi.org/10.1038/nphoton.2016.51}
}

@article{liorni2021quantum,
  author       = {Carlo Liorni and Hermann Kampermann and Dagmar Bru{\ss}},
  title        = {Quantum repeaters in space},
  journal      = {New Journal of Physics},
  volume       = {23},
  number       = {5},
  pages        = {053021},
  year         = {2021},
  publisher    = {IOP Publishing},
  doi          = {10.1088/1367-2630/abfa63},
  url          = {https://doi.org/10.1088/1367-2630/abfa63}
}

@article{simon2007quantum,
  author       = {Christoph Simon and Hugues de Riedmatten and Mikael Afzelius and Nicolas Sangouard and Hugo Zbinden and Nicolas Gisin},
  title        = {Quantum Repeaters with Photon Pair Sources and Multimode Memories},
  journal      = {Physical Review Letters},
  volume       = {98},
  number       = {19},
  pages        = {190503},
  year         = {2007},
  publisher    = {American Physical Society},
  doi          = {10.1103/PhysRevLett.98.190503},
  url          = {https://doi.org/10.1103/PhysRevLett.98.190503}
}

@article{ortu2022multimode,
  author       = {Antonio Ortu and Jelena V. Rakonjac and Adrian Holz{\"a}pfel and Alessandro Seri and Samuele Grandi and Margherita Mazzera and Hugues de Riedmatten and Mikael Afzelius},
  title        = {Multimode capacity of atomic-frequency comb quantum memories},
  journal      = {Quantum Science and Technology},
  volume       = {7},
  number       = {3},
  pages        = {035024},
  year         = {2022},
  publisher    = {IOP Publishing},
  doi          = {10.1088/2058-9565/ac73b0},
  url          = {https://doi.org/10.1088/2058-9565/ac73b0}
}

@article{Knaut2024,
  author = "C. M. Knaut and A. Suleymanzade and Y. C. Wei and others",
  title = "Entanglement of nanophotonic quantum memory nodes in a telecom network",
  journal = "Nature",
  volume = "629",
  pages = "573--578",
  year = "2024",
  doi = "10.1038/s41586-024-07252-z",
  url = "https://doi.org/10.1038/s41586-024-07252-z",
  issue = "16 May 2024"
}

@article{Aspelmeyer2018,
  author = "Markus Aspelmeyer and Thomas Jennewein and Martin Pfennigbauer and Walter Leeb and Anton Zeilinger",
  title = "Long-Distance Quantum Communication with Entangled Photons using Satellites",
  journal = "IEEE Journal of Selected Topics in Quantum Electronics",
  year = "2018",
  doi = "10.1109/JSTQE.2003.820918",
  url = "https://doi.org/10.48550/arXiv.quant-ph/0305105",
}

@article{deForgesdeParny2023,
  author = "L. de Forges de Parny and O. Alibart and J. Debaud and others",
  title = "Satellite-based quantum information networks: use cases, architecture, and roadmap",
  journal = "Communications Physics",
  volume = "6",
  pages = "12",
  year = "2023",
  doi = "10.1038/s42005-022-01123-7",
  url = "https://doi.org/10.1038/s42005-022-01123-7",
  published = "16 January 2023"
}

@misc{rotherham2024advancing,
  author       = {Eugene Rotherham and Connor Casey and Eva Fernandez Rodriguez and Karen Wendy Vidaurre Torrez and Maren Mashor and Isaac Pike},
  title        = {Advancing Free-Space Optical Communication System Architecture: Performance Analysis of Varied Optical Ground Station Network Configurations},
  year         = {2024},
  doi          = {10.48550/arXiv.2410.23470},
  url          = {https://arxiv.org/abs/2410.23470}
}

@article{Briegel1998,
  author       = {H.-J. Briegel and W. Dür and J. I. Cirac and P. Zoller},
  title        = {Quantum Repeaters: The Role of Imperfect Local Operations in Quantum Communication},
  journal      = {Physical Review Letters},
  volume       = {81},
  number       = {26},
  pages        = {5932--5935},
  year         = {1998},
  doi          = {10.1103/PhysRevLett.81.5932},
  url          = {https://doi.org/10.1103/PhysRevLett.81.5932},
  publisher    = {American Physical Society}
}

@article{Kalachev2023QuantumRepeaters,
  author = {A. Kalachev},
  title = {Quantum Repeaters: Current Developments and Prospects},
  journal = {Bulletin of the Lebedev Physics Institute},
  volume = {50},
  number = {Suppl 12},
  pages = {S1312--S1329},
  year = {2023},
  doi = {10.3103/S1068335623602212},
  url = {https://doi.org/10.3103/S1068335623602212},
}

@article{zangi2023entanglement,
  author = {S. M. Zangi and Chitra Shukla and Atta Ur Rahman and Bo Zheng},
  title = {Entanglement Swapping and Swapped Entanglement},
  journal = {Entropy},
  volume = {25},
  number = {3},
  pages = {415},
  year = {2023},
  doi = {10.3390/e25030415},
  url = {https://doi.org/10.48550/arXiv.2212.03413},
  }

@article{childs2023streaming,
  author = {Andrew M. Childs and Honghao Fu and Debbie Leung and Zhi Li and Maris Ozols and Vedang Vyas},
  title = {Streaming Quantum State Purification},
  journal = {arXiv preprint},
  volume = {arXiv:2309.16387},
  year = {2023},
  doi = {10.48550/arXiv.2309.16387},
  url = {https://doi.org/10.48550/arXiv.2309.16387},
  }

@article{Pirandola2021,
  title = {Satellite quantum communications: Fundamental bounds and practical security},
  author = {Pirandola, Stefano},
  journal = {Physical Review Research},
  volume = {3},
  number = {2},
  pages = {023130},
  year = {2021},
  publisher = {American Physical Society},
  address = {Department of Computer Science, University of York, York YO10 5GH, United Kingdom},
  doi = {10.1103/PhysRevResearch.3.023130},
}

@article{khatri2021spooky,
  author={Khatri, S. and Brady, A. J. and Desporte, R. A. and others},
  title={Spooky action at a global distance: analysis of space-based entanglement distribution for the quantum internet},
  journal={npj Quantum Information},
  volume={7},
  number={4},
  year={2021},
  doi={10.1038/s41534-020-00327-5}
}

@inproceedings{Mariani2023CosmicRaysQubits,
  author = {A. Mariani and others},
  title = {Mitigation of Cosmic Rays-Induced Errors in Superconducting Quantum Processors},
  booktitle = {2023 IEEE International Conference on Quantum Computing and Engineering (QCE)},
  address = {Bellevue, WA, USA},
  pages = {1389--1393},
  year = {2023},
  doi = {10.1109/QCE57702.2023.00157},
}

@book{VanMeter2014,
  author = "Rodney Van Meter",
  title = "Quantum Networking",
  edition = "1st",
  publisher = "Wiley-ISTE",
  year = "2014",
  isbn = "978-1848215375"
}

@article{Gundogan2021npjQI,
  author  = {G{\"u}ndo\u{g}an, M. and Sidhu, J. S. and Henderson, V. and others},
  title   = {Proposal for Space-Borne Quantum Memories for Global Quantum Networking},
  journal = {npj Quantum Information},
  year    = {2021},
  volume  = {7},
  pages   = {128},
  doi     = {10.1038/s41534-021-00460-9},
  url     = {https://doi.org/10.1038/s41534-021-00460-9}
}

@article{Abruzzo2014PRA,
  title        = {Measurement-device-independent quantum key distribution with quantum memories},
  author       = {Abruzzo, Silvestre and Kampermann, Hermann and Bru{\ss}, Dagmar},
  journal      = {Physical Review A},
  volume       = {89},
  pages        = {012301},
  year         = {2014},
  doi          = {10.1103/PhysRevA.89.012301}
}

@article{Panayi2014NJP,
  title        = {Memory-assisted measurement-device-independent quantum key distribution},
  author       = {Panayi, Christiana and Razavi, Mohsen and Ma, Xiongfeng and L{\"u}tkenhaus, Norbert},
  journal      = {New Journal of Physics},
  volume       = {16},
  pages        = {043005},
  year         = {2014},
  doi          = {10.1088/1367-2630/16/4/043005}
}

@article{Luong2016APB,
  title        = {Overcoming lossy channel bounds using a single quantum repeater node},
  author       = {Luong, Duc and Jiang, Liang and Kim, Jungsang and L{\"u}tkenhaus, Norbert},
  journal      = {Applied Physics B},
  volume       = {122},
  pages        = {96},
  year         = {2016},
  doi          = {10.1007/s00340-016-6373-4},
  url          = {https://doi.org/10.1007/s00340-016-6373-4}
}

@misc{Bae2025arXiv,
  title        = {Blockwise Post-processing in Satellite-based Quantum Key Distribution},
  author       = {Bae, Minu J. and Panigrahy, Nitish K. and Dhara, Prajit and Hossain, Md Zakir and Krawec, Walter O. and Russell, Alexander and Towsley, Don and Wang, Bing},
  eprint       = {2503.06031},
  archivePrefix= {arXiv},
  primaryClass = {quant-ph},
  year         = {2025},
  doi          = {10.48550/arXiv.2503.06031}
}

@article{BennettBrassard2014TCS,
  title        = {Quantum cryptography: Public key distribution and coin tossing},
  author       = {Bennett, Charles H. and Brassard, Gilles},
  journal      = {Theoretical Computer Science},
  volume       = {560},
  pages        = {7--11},
  year         = {2014},
}

@article{Ekert1991PRL,
  title        = {Quantum cryptography based on Bell’s theorem},
  author       = {Ekert, Artur K.},
  journal      = {Physical Review Letters},
  volume       = {67},
  pages        = {661--663},
  year         = {1991},
  doi          = {10.1103/PhysRevLett.67.661}
}

@article{Pirandola2017FundamentalLimits,
  title        = {Fundamental limits of repeaterless quantum communications},
  author       = {Pirandola, Stefano and Laurenza, Riccardo and Ottaviani, Carlo and Banchi, Leonardo},
  journal      = {Nature Communications},
  volume       = {8},
  pages        = {15043},
  year         = {2017},
  doi          = {10.1038/ncomms15043}
}

@article{Pirandola2021FSOBounds,
  title        = {Limits and security of free-space quantum communications},
  author       = {Pirandola, Stefano},
  journal      = {Physical Review Research},
  volume       = {3},
  pages        = {013279},
  year         = {2021},
  doi          = {10.1103/PhysRevResearch.3.013279}
}

@article{Zheng2022PRXQ040319,
  title        = {Entanglement Distribution with Minimal Memory Requirements Using Time-Bin Photonic Qudits},
  author       = {Zheng, Yunzhe and Sharma, Hemant and Borregaard, Johannes},
  journal      = {PRX Quantum},
  volume       = {3},
  pages        = {040319},
  year         = {2022},
  doi          = {10.1103/PRXQuantum.3.040319}
}

@article{Azuma2023RMP,
  title        = {Quantum repeaters: From quantum networks to the quantum internet},
  author       = {Azuma, Koji and Economou, Sophia E. and Elkouss, David and Hilaire, Paul and Jiang, Liang and Lo, Hoi-Kwong and Tzitrin, Ilan},
  journal      = {Reviews of Modern Physics},
  volume       = {95},
  number       = {4},
  pages        = {045006},
  year         = {2023},
  doi          = {10.1103/RevModPhys.95.045006}
}

@article{Wei2024npjQI,
  title        = {Quantum storage of 1650 modes of single photons at telecom wavelength},
  author       = {Wei, Shang-Hua and Jing, Bo and Zhang, Xiao-Yu and others},
  journal      = {npj Quantum Information},
  volume       = {10},
  pages        = {19},
  year         = {2024},
  doi          = {10.1038/s41534-024-00812-1}
}

@article{Afzelius2009PRA,
  title        = {Multimode quantum memory based on atomic frequency combs},
  author       = {Afzelius, Mikael and Simon, Christoph and de Riedmatten, Hugues and Gisin, Nicolas},
  journal      = {Physical Review A},
  volume       = {79},
  pages        = {052329},
  year         = {2009},
  doi          = {10.1103/PhysRevA.79.052329}
}

@article{LagoRivera2021Nature,
  title        = {Telecom-heralded entanglement between multimode solid-state quantum memories},
  author       = {Lago-Rivera, D. and Grandi, S. and Rakonjac, J. V. and Seri, A. and de Riedmatten, H.},
  journal      = {Nature},
  volume       = {594},
  pages        = {37--40},
  year         = {2021},
  doi          = {10.1038/s41586-021-03481-8}
}

@article{Liu2021Nature,
  title        = {Heralded entanglement distribution between two absorptive quantum memories},
  author       = {Liu, X. and Hu, J. and Li, Z.-F. and Li, X. and Li, P.-Y. and Liang, P.-J. and Zhou, Z.-Q. and Li, C.-F. and Guo, G.-C.},
  journal      = {Nature},
  volume       = {594},
  pages        = {41--45},
  year         = {2021},
  doi          = {10.1038/s41586-021-03505-3}
}

@article{Heinze2013PRL,
  title        = {Stopped light and image storage by electromagnetically induced transparency up to the regime of one minute},
  author       = {Heinze, Georg and Hubrich, Christoph and Halfmann, Thomas},
  journal      = {Physical Review Letters},
  volume       = {111},
  pages        = {033601},
  year         = {2013},
  doi          = {10.1103/PhysRevLett.111.033601}
}

@article{Usmani2012NatPhoton,
  title        = {Heralded quantum entanglement between two crystals},
  author       = {Usmani, I. and Clausen, C. and Bussi{\`e}res, F. and Sangouard, N. and Afzelius, M. and Gisin, N.},
  journal      = {Nature Photonics},
  volume       = {6},
  pages        = {234--237},
  year         = {2012},
  doi          = {10.1038/nphoton.2012.67}
}

@article{Gundogan2013NJP,
  title        = {Coherent storage of temporally multimode light using a spin-wave atomic frequency comb memory},
  author       = {G{\"u}ndo{\u{g}}an, M. and Mazzera, M. and Ledingham, P. M. and Cristiani, M. and de Riedmatten, H.},
  journal      = {New Journal of Physics},
  volume       = {15},
  pages        = {045012},
  year         = {2013},
  doi          = {10.1088/1367-2630/15/4/045012}
}

@article{Seri2019PRL,
  title        = {Quantum storage of frequency-multiplexed heralded single photons},
  author       = {Seri, A. and Lago-Rivera, D. and Lenhard, A. and Corrielli, G. and Osellame, R. and Mazzera, M. and de Riedmatten, H.},
  journal      = {Physical Review Letters},
  volume       = {123},
  pages        = {080502},
  year         = {2019},
  doi          = {10.1103/PhysRevLett.123.080502}
}

@article{Sinclair2014PRL,
  title        = {Spectral multiplexing for scalable quantum photonics using an atomic frequency comb quantum memory and feed-forward control},
  author       = {Sinclair, N. and Saglamyurek, E. and Mallahzadeh, H. and Slater, J. A. and George, M. and Ricken, R. and Hedges, M. P. and Oblak, D. and Simon, C. and Sohler, W. and Tittel, W.},
  journal      = {Physical Review Letters},
  volume       = {113},
  pages        = {053603},
  year         = {2014},
  doi          = {10.1103/PhysRevLett.113.053603}
}

@article{Saglamyurek2016NatComm,
  title        = {A multiplexed light-matter interface for fibre-based quantum networks},
  author       = {Saglamyurek, E. and Grimau Puigibert, M. and Zhou, Q. and Giner, L. and Marsili, F. and Verma, V. B. and Nam, S. W. and Oesterling, L. and Nippa, D. and Oblak, D. and Tittel, W.},
  journal      = {Nature Communications},
  volume       = {7},
  pages        = {11202},
  year         = {2016},
  doi          = {10.1038/ncomms11202}
}

@article{Yang2018NatComm,
  title        = {Multiplexed storage and real-time manipulation based on a multiple degree-of-freedom quantum memory},
  author       = {Yang, T.-S. and Zhou, Z.-Q. and Hua, Y.-L. and Liu, X. and Li, Z.-F. and Li, P.-Y. and Ma, Y. and Liu, C. and Liang, P.-J. and Li, X. and others},
  journal      = {Nature Communications},
  volume       = {9},
  pages        = {3407},
  year         = {2018},
  doi          = {10.1038/s41467-018-05669-5}
}

@article{Saglamyurek2011Nature,
  title        = {Broadband waveguide quantum memory for entangled photons},
  author       = {Saglamyurek, E. and Sinclair, N. and Jin, J. and Slater, J. A. and Oblak, D. and Bussi{\`e}res, F. and George, M. and Ricken, R. and Sohler, W. and Tittel, W.},
  journal      = {Nature},
  volume       = {469},
  pages        = {512--515},
  year         = {2011},
  doi          = {10.1038/nature09719}
}

@article{Marzban2015PRL,
  title        = {Observation of photon echoes from evanescently coupled rare-earth ions in a planar waveguide},
  author       = {Marzban, S. and Bartholomew, J. G. and Madden, S. and Vu, K. and Sellars, M. J.},
  journal      = {Physical Review Letters},
  volume       = {115},
  pages        = {013601},
  year         = {2015},
  doi          = {10.1103/PhysRevLett.115.013601}
}

@article{Corrielli2016PRApplied,
  title        = {Integrated optical memory based on laser-written waveguides},
  author       = {Corrielli, G. and Seri, A. and Mazzera, M. and Osellame, R. and de Riedmatten, H.},
  journal      = {Physical Review Applied},
  volume       = {5},
  pages        = {054013},
  year         = {2016},
  doi          = {10.1103/PhysRevApplied.5.054013}
}

@article{Zhong2017Science,
  title        = {Nanophotonic rare-earth quantum memory with optically controlled retrieval},
  author       = {Zhong, Tian and Kindem, Jonathan M. and Bartholomew, John G. and Rochman, Jonathan and Craiciu, Iulia and Miyazono, Evan and Bettinelli, Marco and Cavalli, Enrico and Verma, Varun and Nam, Sae Woo and Marsili, Francesco and Shaw, Matthew D. and Beyer, Alexander D. and Faraon, Andrei},
  journal      = {Science},
  volume       = {357},
  number       = {6358},
  pages        = {1392--1395},
  year         = {2017},
  doi          = {10.1126/science.aan5959}
}

@article{Dibos2018PRL,
  title        = {Atomic source of single photons in the telecom band},
  author       = {Dibos, Amos M. and Raha, Mousol and Phenicie, Cody M. and Thompson, Jeff D.},
  journal      = {Physical Review Letters},
  volume       = {120},
  pages        = {243601},
  year         = {2018},
  doi          = {10.1103/PhysRevLett.120.243601}
}

@article{Jobez2015PRL,
  title        = {Coherent spin control at the quantum level in an ensemble-based optical memory},
  author       = {Jobez, P. and Laplane, C. and Timoney, N. and Gisin, N. and Ferrier, A. and Goldner, P. and Afzelius, M.},
  journal      = {Physical Review Letters},
  volume       = {114},
  pages        = {230502},
  year         = {2015},
  doi          = {10.1103/PhysRevLett.114.230502}
}

@article{Gundogan2015PRL,
  title        = {Solid state spin-wave quantum memory for time-bin qubits},
  author       = {G{\"u}ndo{\u{g}}an, M. and Ledingham, P. M. and Kutluer, K. and Mazzera, M. and de Riedmatten, H.},
  journal      = {Physical Review Letters},
  volume       = {114},
  pages        = {230501},
  year         = {2015},
  doi          = {10.1103/PhysRevLett.114.230501}
}

@article{Businger2020PRL,
  title        = {Optical spin-wave storage in a solid-state hybridized electron-nuclear spin ensemble},
  author       = {Businger, M. and others},
  journal      = {Physical Review Letters},
  volume       = {124},
  pages        = {053606},
  year         = {2020},
  doi          = {10.1103/PhysRevLett.124.053606}
}

@article{You2018Opex,
  title        = {Superconducting nanowire single-photon detection system for space applications},
  author       = {You, Lixing and others},
  journal      = {Optics Express},
  volume       = {26},
  pages        = {2965--2971},
  year         = {2018},
  doi          = {10.1364/OE.26.002965}
}

@article{Cho2016Optica,
  title        = {Highly efficient optical quantum memory with long coherence time in cold atoms},
  author       = {Cho, Y.-W. and Campbell, G. T. and Everett, J. L. and Bernu, J. and Higginbottom, D. B. and Cao, M. T. and Geng, J. and Robins, N. P. and Lam, P. K. and Buchler, B. C.},
  journal      = {Optica},
  volume       = {3},
  pages        = {100--107},
  year         = {2016},
  doi          = {10.1364/OPTICA.3.000100}
}

@article{Dudin2013PRA,
  title        = {Light storage on the time scale of a minute},
  author       = {Dudin, Y. O. and Li, L. and Kuzmich, A.},
  journal      = {Physical Review A},
  volume       = {87},
  pages        = {031801},
  year         = {2013},
  doi          = {10.1103/PhysRevA.87.031801}
}

@article{Heller2020PRL,
  title        = {Cold-atom temporally multiplexed quantum memory with cavity-enhanced noise suppression},
  author       = {Heller, L. and Farrera, P. and Heinze, G. and de Riedmatten, H.},
  journal      = {Physical Review Letters},
  volume       = {124},
  pages        = {210504},
  year         = {2020},
  doi          = {10.1103/PhysRevLett.124.210504}
}

@article{Lan2009Opex,
  title        = {A multiplexed quantum memory},
  author       = {Lan, S.-Y. and Radnaev, A. G. and Collins, O. A. and Matsukevich, D. N. and Kennedy, T. A. B. and Kuzmich, A.},
  journal      = {Optics Express},
  volume       = {17},
  pages        = {13639--13645},
  year         = {2009},
  doi          = {10.1364/OE.17.013639}
}

@article{Pu2017NatComm,
  title        = {Experimental realization of a multiplexed quantum memory with 225 individually accessible memory cells},
  author       = {Pu, Y.-F. and Jiang, N. and Chang, W. and Yang, H.-X. and Li, C. and Duan, L.-M.},
  journal      = {Nature Communications},
  volume       = {8},
  pages        = {15359},
  year         = {2017},
  doi          = {10.1038/ncomms15359}
}

@article{Langlois2018PRApplied,
  title        = {Compact cold-atom clock for onboard timebase: Tests in reduced gravity},
  author       = {Langlois, M. and De Sarlo, L. and Holleville, D. and Dimarcq, N. and Schaff, J.-F. and Bernon, S.},
  journal      = {Physical Review Applied},
  volume       = {10},
  pages        = {064007},
  year         = {2018},
  doi          = {10.1103/PhysRevApplied.10.064007}
}

@article{Barrett2016NatComm,
  title        = {Dual matter-wave inertial sensors in weightlessness},
  author       = {Barrett, B. and Antoni-Micollier, L. and Chichet, L. and Battelier, B. and L{\'e}v{\`e}que, T. and Landragin, A. and Bouyer, P.},
  journal      = {Nature Communications},
  volume       = {7},
  pages        = {13786},
  year         = {2016},
  doi          = {10.1038/ncomms13786}
}

@article{Muntinga2013PRL,
  title        = {Interferometry with Bose--Einstein condensates in microgravity},
  author       = {M{\"u}ntinga, H. and Ahlers, H. and Becker, D. and others},
  journal      = {Physical Review Letters},
  volume       = {110},
  pages        = {093602},
  year         = {2013},
  doi          = {10.1103/PhysRevLett.110.093602}
}

@article{Deppner2021PRL,
  title        = {Collective-mode enhanced matter-wave optics},
  author       = {Deppner, C. and Herr, W. and Cornelius, M. and others},
  journal      = {Physical Review Letters},
  volume       = {127},
  pages        = {100401},
  year         = {2021},
  doi          = {10.1103/PhysRevLett.127.100401}
}

@article{Becker2018Nature,
  title        = {Space-borne Bose--Einstein condensation for precision interferometry},
  author       = {Becker, Dennis and Lachmann, M. D. and Seidel, S. T. and others},
  journal      = {Nature},
  volume       = {562},
  pages        = {391--395},
  year         = {2018},
  doi          = {10.1038/s41586-018-0605-1}
}

@article{Lachmann2021NatComm,
  title        = {Ultracold atom interferometry in space},
  author       = {Lachmann, M. D. and Ahlers, H. and Becker, D. and others},
  journal      = {Nature Communications},
  volume       = {12},
  pages        = {1317},
  year         = {2021},
  doi          = {10.1038/s41467-021-21628-z},
  url          = {https://doi.org/10.1038/s41467-021-21628-z}
}

@article{Liu2018NatComm,
  title        = {In-orbit operation of an atomic clock based on laser-cooled $^{87}$Rb atoms},
  author       = {Liu, L. and others},
  journal      = {Nature Communications},
  volume       = {9},
  pages        = {2760},
  year         = {2018},
  doi          = {10.1038/s41467-018-05219-z}
}

@article{Aveline2020Nature,
  title        = {Observation of Bose--Einstein condensates in an Earth-orbiting research lab},
  author       = {Aveline, D. C. and Williams, J. R. and Elliott, E. R. and others},
  journal      = {Nature},
  volume       = {582},
  pages        = {193--197},
  year         = {2020},
  doi          = {10.1038/s41586-020-2346-1}
}

@article{Laurent2015CRPhys,
  title        = {The ACES/PHARAO space mission},
  author       = {Laurent, P. and Massonnet, D. and Cacciapuoti, L. and Salomon, C.},
  journal      = {Comptes Rendus Physique},
  volume       = {16},
  number       = {5},
  pages        = {540--552},
  year         = {2015},
  doi          = {10.1016/j.crhy.2015.05.002},
  url          = {https://doi.org/10.1016/j.crhy.2015.05.002}
}

@article{Frye2021EPJQT,
  title        = {The Bose--Einstein Condensate and Cold Atom Laboratory},
  author       = {Frye, K. and Abend, S. and Bartosch, W. and others},
  journal      = {EPJ Quantum Technology},
  volume       = {8},
  pages        = {1},
  year         = {2021},
  doi          = {10.1140/epjqt/s40507-020-00090-8}
}

@article{Devani2020CEAS,
  title        = {Gravity sensing: cold atom trap onboard a 6U CubeSat},
  author       = {Devani, D. and Maddox, S. and Renshaw, R. and others},
  journal      = {CEAS Space Journal},
  volume       = {12},
  pages        = {539--549},
  year         = {2020},
  doi          = {10.1007/s12567-020-00326-4}
}

@article{ElNeaj2020EPJQT,
  title        = {AEDGE: Atomic Experiment for Dark Matter and Gravity Exploration in Space},
  author       = {El-Neaj, Y. A. and Alpigiani, C. and Amairi-Pyka, S. and others},
  journal      = {EPJ Quantum Technology},
  volume       = {7},
  pages        = {6},
  year         = {2020},
  doi          = {10.1140/epjqt/s40507-020-0080-0}
}

@article{Strangfeld2021JOSAB,
  title        = {Prototype of a compact rubidium-based optical frequency reference for operation on nanosatellites},
  author       = {Strangfeld, S. and Kanthak, S. and Schiemangk, M. and Wiegand, I. and Wicht, A. and Ling, A. and Krutzik, M.},
  journal      = {JOSA B},
  volume       = {38},
  pages        = {1885--1891},
  year         = {2021},
  doi          = {10.1364/JOSAB.420875},
  url          = {https://doi.org/10.1364/JOSAB.420875
}
}

@article{Fragner2008Science,
  title        = {Resolving vacuum fluctuations of an electrical field in a superconducting circuit},
  author       = {Fragner, Andreas and G{\"o}ppl, Markus and Fink, Johannes M. and Blais, Alexandre and Esteve, Daniel and Wallraff, Andreas},
  journal      = {Science},
  volume       = {322},
  number       = {5906},
  pages        = {1357--1360},
  year         = {2008},
  doi          = {10.1126/science.1164482}
}

@article{Fowler2024PRXQ,
  title        = {Spectroscopic Measurements and Models of Energy Deposition in the Substrate of Quantum Circuits by Natural Ionizing Radiation},
  author       = {Fowler, Joseph W. and Shank, Benjamin R. and Brown, Wendy C. and others},
  journal      = {PRX Quantum},
  volume       = {5},
  pages        = {040323},
  year         = {2024},
  doi          = {10.1103/PRXQuantum.5.040323}
}

@article{Katz2021SciAdv,
  title        = {Coupling light to a nuclear spin gas with a two-photon resonance},
  author       = {Katz, Or and Peleg, Yuval and Shaham, Roy and Firstenberg, Ofer},
  journal      = {Science Advances},
  volume       = {7},
  number       = {23},
  pages        = {eabe9164},
  year         = {2021},
  doi          = {10.1126/sciadv.abe9164}
}

@article{Katz2022PRA,
  title        = {Optical quantum memory for noble-gas spins based on spin-exchange collisions},
  author       = {Katz, Or and Shaham, Roy and Firstenberg, Ofer},
  journal      = {Physical Review A},
  volume       = {105},
  pages        = {042606},
  year         = {2022},
  doi          = {10.1103/PhysRevA.105.042606}
}

@misc{Xing2024arXivCOTS,
  title        = {Computing assigned to satellite: Leveraging COTS hardware at LEO},
  author       = {Xing, Weilin and Li, Zhen and others},
  year         = {2024},
  eprint       = {2401.03435},
}

@article{ogCRYOpaper,
  title = {Noncryogenic Quantum Repeaters with hot Hybrid Alkali-Noble Gases},
  author = {Ji, Jia-Wei and Asadi, Faezeh Kimiaee and Heshami, Khabat and Simon, Christoph},
  journal = {Phys. Rev. Appl.},
  volume = {19},
  issue = {5},
  pages = {054063},
  numpages = {12},
  year = {2023},
  month = {May},
  publisher = {American Physical Society},
  doi = {10.1103/PhysRevApplied.19.054063},
  url = {https://link.aps.org/doi/10.1103/PhysRevApplied.19.054063}
}

@article{Ortu2022QST_AFCmultimode,
  author    = {Ortu, Antonio and Rakonjac, Jelena V. and Holz{\"a}pfel, Adrian and Seri, Alessandro and Grandi, Samuele and Mazzera, Margherita and {de Riedmatten}, Hugues and Afzelius, Mikael},
  title     = {Multimode capacity of atomic-frequency comb quantum memories},
  journal   = {Quantum Science and Technology},
  year      = {2022},
  volume    = {7},
  number    = {3},
  pages     = {035024},
  month     = {jun},
  publisher = {IOP Publishing},
  doi       = {10.1088/2058-9565/ac73b0},
  url       = {https://doi.org/10.1088/2058-9565/ac73b0}
}

@inproceedings{Dassie2023Doppler,
  author    = {Dassi{\'e}, Manuele and Giorgi, Gabriele and Dominguez, Pablo Nahuel and Bl{\"u}mel, Ludwig and Gohle, Christoph},
  title     = {Doppler Compensation for Optical Inter-Satellite and Satellite-to-Ground Frequency Transfer},
  booktitle = {Proceedings of the 36th International Technical Meeting of the Satellite Division of The Institute of Navigation (ION GNSS+ 2023)},
  address   = {Denver, Colorado},
  month     = sep,
  year      = {2023},
  pages     = {3190--3204},
  doi       = {10.33012/2023.19257},
  url       = {https://doi.org/10.33012/2023.19257},
  organization = {The Institute of Navigation}
}

@article{Ahmadi2024QUICK3,
  author  = {Ahmadi, Najme and Schwertfeger, Sven and Werner, Philipp and Wiese, Lukas and Lester, Joseph and {Da Ros}, Elisa and Krause, Josefine and Ritter, Sebastian and Abasifard, Mostafa and Cholsuk, Chanaprom and Kr{\"a}mer, Ria G. and Atzeni, Simone and G{\"u}ndo\u{g}an, Mustafa and Sachidananda, Subash and Pardo, Daniel and Nolte, Stefan and Lohrmann, Alexander and Ling, Alexander and Bartholom{\"a}us, Julian and Corrielli, Giacomo and Krutzik, Markus and Vogl, Tobias},
  title   = {QUICK$^{3}$: Design of a Satellite-Based Quantum Light Source for Quantum Communication and Extended Physical Theory Tests in Space},
  journal = {Advanced Quantum Technologies},
  year    = {2024},
  volume  = {7},
  pages   = {2300343},
  doi     = {10.1002/qute.202300343},
  url     = {https://doi.org/10.1002/qute.202300343}
}

@article{Bedington2017Progress,
  author  = {Bedington, Robert and Arrazola, Juan Miguel and Ling, Alexander},
  title   = {Progress in Satellite Quantum Key Distribution},
  journal = {npj Quantum Information},
  year    = {2017},
  volume  = {3},
  pages   = {30},
  doi     = {10.1038/s41534-017-0031-5},
  url     = {https://doi.org/10.1038/s41534-017-0031-5}
}

@article{Elliott2018CAL,
  author  = {Elliott, E. R. and Krutzik, M. C. and Williams, J. R. and others},
  title   = {NASA’s Cold Atom Lab (CAL): System Development and Ground Test Status},
  journal = {npj Microgravity},
  year    = {2018},
  volume  = {4},
  pages   = {16},
  doi     = {10.1038/s41526-018-0049-9},
  url     = {https://doi.org/10.1038/s41526-018-0049-9}
}

@techreport{S3VI2025ThermalControl,
  author       = {{NASA Small Spacecraft Systems Virtual Institute (S3VI)}},
  title        = {State-of-the-Art of Small Spacecraft Technology: Chapter 7 — Thermal Control (2024 Edition)},
  institution  = {National Aeronautics and Space Administration},
  address      = {Moffett Field, CA, USA},
  year         = {2025},
  type         = {Technical Report}
}

@article{Wang2022MicroFabricatedVaporCells,
  author    = {Wang, X. and Ye, M. and Lu, F. and Mao, Y. and Tian, H. and Li, J.},
  title     = {Recent Progress on Micro-Fabricated Alkali Metal Vapor Cells},
  journal   = {Biosensors},
  year      = {2022},
  month     = mar,
  volume    = {12},
  number    = {3},
  pages     = {165},
  doi       = {10.3390/bios12030165},
  url       = {https://doi.org/10.3390/bios12030165},
  pmid      = {35323435},
  pmcid     = {PMC8946820},
  publisher = {MDPI}
}

@article{Zhang2018LargeScaleQKD,
  author  = {Zhang, Qiang and Xu, Feihu and Chen, Yu-Ao and Peng, Cheng-Zhi and Pan, Jian-Wei},
  title   = {Large Scale Quantum Key Distribution: Challenges and Solutions {[Invited]}},
  journal = {Optics Express},
  year    = {2018},
  volume  = {26},
  number  = {18},
  pages   = {24260--24273},
  doi     = {10.1364/OE.26.024260},
  url     = {https://doi.org/10.1364/OE.26.024260}
}

@article{Wallnoefer2022Simulating,
  author  = {Walln{\"o}fer, J. and Hahn, F. and G{\"u}ndo\u{g}an, M. and others},
  title   = {Simulating quantum repeater strategies for multiple satellites},
  journal = {Communications Physics},
  year    = {2022},
  volume  = {5},
  pages   = {169},
  doi     = {10.1038/s42005-022-00945-9},
  url     = {https://doi.org/10.1038/s42005-022-00945-9}
}

@article{Ma2007EntangledSourcesQKD,
  author  = {Ma, Xiongfeng and Fung, C.-H. F. and Lo, H.-K.},
  title   = {Quantum key distribution with entangled photon sources},
  journal = {Physical Review A},
  volume  = {76},
  pages   = {012307},
  year    = {2007},
  doi     = {10.1103/PhysRevA.76.012307}
}

@article{Trenyi2020MQMBeatingDirect,
  author  = {Tr{\'e}nyi, R. and L{\"u}tkenhaus, Norbert},
  title   = {Beating direct transmission bounds for quantum key distribution with a multiple quantum memory station},
  journal = {Physical Review A},
  volume  = {101},
  pages   = {012325},
  year    = {2020},
  doi     = {10.1103/PhysRevA.101.012325}
}

@article{Katz2022QuantumInterface,
  author       = {Katz, Or and Shaham, Roy and Firstenberg, Ofer},
  title        = {Quantum interface for noble-gas spins based on spin-exchange collisions},
  journal      = {PRX Quantum},
  year         = {2022},
  volume       = {3},
  pages        = {010305},
  doi          = {10.1103/PRXQuantum.3.010305},
}

@article{Beysens2022TransportMicrogravity,
  author  = {Beysens, D.},
  title   = {Editorial: Transport phenomena in microgravity},
  journal = {Frontiers in Space Technologies},
  year    = {2022},
  volume  = {3},
  pages   = {1092802},
  doi     = {10.3389/frspt.2022.1092802},
  url     = {https://doi.org/10.3389/frspt.2022.1092802}
}

@techreport{Niederhaus2008IVMixing,
  author      = {Niederhaus, Charles E. and Miller, Fletcher J.},
  title       = {Intravenous Fluid Mixing in Normal Gravity, Partial Gravity, and Microgravity: Down-Selection of Mixing Methods},
  institution = {NASA Glenn Research Center},
  address     = {Cleveland, Ohio},
  year        = {2008},
  number      = {NASA/TM-2008-215000},
  type        = {NASA Technical Memorandum},
  url         = {https://ntrs.nasa.gov/citations/20080012740}
}

@article{Selvadurai2022PassiveThermal,
  author  = {Selvadurai, Shanmugasundaram and Chandran, Amal and Valentini, David and Lamprecht, Bret},
  title   = {Passive Thermal Control Design Methods, Analysis, Comparison, and Evaluation for Micro and Nanosatellites Carrying Infrared Imager},
  journal = {Applied Sciences},
  year    = {2022},
  volume  = {12},
  number  = {6},
  pages   = {2858},
  doi     = {10.3390/app12062858},
  url     = {https://doi.org/10.3390/app12062858}
}

@article{Binder1975SatelliteAnomalies,
  author  = {Binder, D. and Smith, E. C. and Holman, A. B.},
  title   = {Satellite Anomalies from Galactic Cosmic Rays},
  journal = {IEEE Transactions on Nuclear Science},
  year    = {1975},
  volume  = {22},
  number  = {6},
  pages   = {2675--2680},
  doi     = {10.1109/TNS.1975.4328188},
  issn    = {0018-9499}
}

@article{Daneshvar2021MultilayerShield,
  author  = {Daneshvar, H. and Milan, K. G. and Sadr, A. and Ghasemi, M. and Abrishami, S.},
  title   = {Multilayer radiation shield for satellite electronic components protection},
  journal = {Scientific Reports},
  year    = {2021},
  volume  = {11},
  number  = {1},
  pages   = {20657},
  doi     = {10.1038/s41598-021-99739-2},
  url     = {https://doi.org/10.1038/s41598-021-99739-2}
}

@article{conductiveCoating,
  title = {Measurement of dc and ac Electric Fields inside an Atomic Vapor Cell with Wall-Integrated Electrodes},
  author = {Ma, Lu and Viray, Michael A. and Anderson, David A. and Raithel, Georg},
  journal = {Phys. Rev. Appl.},
  volume = {18},
  issue = {2},
  pages = {024001},
  numpages = {9},
  year = {2022},
  month = {Aug},
  publisher = {American Physical Society},
  doi = {10.1103/PhysRevApplied.18.024001},
  url = {https://link.aps.org/doi/10.1103/PhysRevApplied.18.024001}
}

@article{Gentile2017OpticallyPolarizedHe3,
  title   = {Optically polarized $^3$He},
  author  = {Gentile, T. R. and Nacher, P. J. and Saam, B. and Walker, T. G.},
  journal = {Reviews of Modern Physics},
  volume  = {89},
  number  = {4},
  pages   = {045004},
  year    = {2017},
  doi     = {10.1103/RevModPhys.89.045004},
  publisher = {American Physical Society}
}

@article{Wu2021WallInteractionsSpinPolarizedAtoms,
  title     = {Wall interactions of spin-polarized atoms},
  author    = {Wu, Zhen},
  journal   = {Reviews of Modern Physics},
  volume    = {93},
  number    = {3},
  pages     = {035006},
  year      = {2021},
  doi       = {10.1103/RevModPhys.93.035006},
  publisher = {American Physical Society}
}

@article{Chang2024WallCollisionEffect,
  title     = {Wall-collision effect on optically polarized atoms in small and hot vapor cells},
  author    = {Chang, Yue and Qin, Jie},
  journal   = {Physical Review A},
  volume    = {109},
  number    = {2},
  pages     = {023113},
  year      = {2024},
  doi       = {10.1103/PhysRevA.109.023113},
  publisher = {American Physical Society}
}

@article{Shaham2022StrongCoupling,
  title   = {Strong coupling of alkali-metal spins to noble-gas spins with an hour-long coherence time},
  author  = {Shaham, Roy and Katz, Or and Firstenberg, Ofer},
  journal = {Nature Physics},
  volume  = {18},
  number  = {5},
  pages   = {506--510},
  year    = {2022},
  doi     = {10.1038/s41567-022-01535-w},
  publisher = {Nature Publishing Group}
}

@article{Gensterblum2015GasTransportShale,
  title   = {Gas transport and storage capacity in shale gas reservoirs -- A review. Part A: Transport processes},
  author  = {Gensterblum, Yves and Ghanizadeh, Amin and Cuss, Robert J. and Amann-Hildenbrand, Alexandra and Krooss, Bernhard M. and Clarkson, Christopher R. and Harrington, John F. and Zoback, Mark D.},
  journal = {Journal of Unconventional Oil and Gas Resources},
  volume  = {12},
  pages   = {87--122},
  year    = {2015},
  doi     = {10.1016/j.juogr.2015.08.001},
  publisher = {Elsevier}
}

\appendix
\clearpage

\section{Probability of establishing a link}
\label{app:link-prob}

We model each elementary connection as four statistically independent processes, following the Supplementary Information in Ref. \cite{Wallnoefer2022Simulating}. See the full derivation in the cited references, but the main points are summarized below. 

\begin{enumerate}
\item \textbf{Memory Efficiency \(\eta_{\mathrm{mem}}\).}
Probability to convert an incoming photon into a stationary qubit and retrieve it later. From the results of this study, we adopt \(\eta_{\mathrm{mem}}=0.74\).

\item \textbf{Detector Efficiency \(\eta_{\mathrm{det}}\).}
Single-photon detection at the readout terminal.  
Given demonstrated SNSPD performance, we adopt \(\eta_{\mathrm{det}}=0.7\) for ground stations which is consistent with \cite{Wallnoefer2022Simulating}.

\item \textbf{Atmospheric Transmission \(\eta_{\mathrm{atm}}(\theta)\).}
Beer–Lambert scaling with elevation angle \(\theta\):
\begin{equation}
\eta_{\mathrm{atm}}(\theta)
= \bigl(\eta_{\mathrm{atm}}^{z}\bigr)^{1/\sin\theta},
\qquad
\eta_{\mathrm{atm}}^{z}=0.8 \;\; (\lambda\!\approx\!795~\mathrm{nm}),
\end{equation}
appropriate for clear-sky mid-latitude conditions \cite{khatri2021spooky,Gundogan2021npjQI}.

\item \textbf{Diffraction and Pointing Loss \(\eta_{\mathrm{dif}}(l)\).}
Beam spreading and residual pointing jitter over slant range \(l\).  
A paraxial, diffraction-limited Gaussian emitted with waist \(w_0\) obeys
\begin{align}
I(r,0) &= I_0\,e^{-2r^2/w_0^2},\\
\theta_d &= \frac{\lambda}{\pi w_0},\\
w(z) &= w_0\sqrt{1+\Big(\tfrac{\theta_d}{w_0}z\Big)^2},\\
I(r,z) &= I_0\Big(\tfrac{w_0}{w(z)}\Big)^2 e^{-2r^2/w(z)^2}.
\end{align}
Pointing jitter is modeled as a 2D Gaussian with standard deviation \(\sigma_p\):
\begin{equation}
g_p(\boldsymbol r)
= \frac{1}{2\pi\sigma_p^2}\exp\!\Big(-\tfrac{\|\boldsymbol r\|^2}{2\sigma_p^2}\Big).
\end{equation}
The far-field intensity is the 2D convolution (compact two-column form):
\begin{align}
\tilde I(\boldsymbol r,z)
&= \!\int_{\mathbb{R}^2}\! d^2\boldsymbol\rho\;
   I(\boldsymbol\rho,z)\,g_p(\boldsymbol r-\boldsymbol\rho)
\label{eq:conv2d}\\[-2pt]
&= (I * g_p)(\boldsymbol r;z).
\nonumber
\end{align}
Collected power through a circular aperture of radius \(R_{\mathrm{rec}}\) at range \(l\) is
\begin{align}
P(l)
&= \int_{0}^{R_{\mathrm{rec}}}\!\rho\,d\rho
    \int_{0}^{2\pi}\!d\varphi\;\tilde I(\rho,l),
\label{eq:Pl}\\
\eta_{\mathrm{dif}}(l) &= \frac{P(l)}{P_0}.
\end{align}
For well-stabilized platforms with \(\sigma_p\!\lesssim\!3~\mu\mathrm{rad}\), receiver-side mispointing is typically negligible because the telescope field of view \(\gg \theta_d,\sigma_p\).
\end{enumerate}

\noindent\textbf{Single-shot success probability.}
The overall probability for an elementary link is thus
\begin{equation}
\eta
= \eta_{\mathrm{mem}}\,
  \eta_{\mathrm{det}}\,
  \eta_{\mathrm{atm}}(\theta)\,
  \eta_{\mathrm{dif}}(l).
\label{eq:eta_total}
\end{equation}

\subsection{Geometric Dependence}
Elevation \(\theta\) and slant range \(l\) from ground-track separation \(L_g\) and satellite altitude \(h\):
\begin{align}
l^2
&= R_E^2 + (R_E + h)^2
   - 2R_E(R_E + h)\cos\!\Big(\tfrac{L_g}{R_E}\Big),
\label{eq:slant}\\
\theta
&= \frac{\pi}{2} - \frac{L_g}{R_E}
   - \arcsin\!\Big[\frac{R_E}{l}\sin\!\Big(\tfrac{L_g}{R_E}\Big)\Big],
\label{eq:elev}
\end{align}
with \(R_E=6{,}371~\mathrm{km}\).%

\section{Simulation‑Parameter Tables}
\label{app:parameters}

\begin{table}[H]
  \centering
  \caption{Communication‑Link Parameters.}
  \label{tab:comm-link}
  \begin{tabular*}{\linewidth}{@{\extracolsep{\fill}}ll}
    \toprule
    Parameter & Value \\ \midrule
    Operating wavelength $\lambda$              & $795\,\mathrm{nm}$\\
    Receive aperture $D_{r}$                    & $1.0\,\mathrm{m}$\\
    Beam‑divergence half‑angle $\theta_{d}$     & $3\,\mu\mathrm{rad}$\\
    Pointing‑jitter rms $\theta_{\mathrm{err}}$ & $1\,\mu\mathrm{rad}$\\
    Minimum elevation angle                     & $20^{\circ}$\\
    \bottomrule
  \end{tabular*}
\end{table}


\begin{table}[H]
  \centering
  \caption{Quantum Satellite Parameters.}
  \label{tab:qr-params}
  \begin{tabular*}{\linewidth}{@{\extracolsep{\fill}}ll}
    \toprule
    Parameter & Value \\ \midrule
    Source repetition rate $R_{\mathrm{rep}}$   & $90\,\mathrm{MHz}$\\
    Memory efficiency $\eta_{\mathrm{mem}}$     & $0.74$\\
    Detector efficiency $\eta_{\mathrm{det}}$   & $0.70$\\
    Bell‑state‑measurement efficiency           & $0.50$\\
    Memory modes                                & $112$\\
    Entangled‑photon source efficiency          & $0.20$\\
    QND‑measurement efficiency                  & $0.80$\\
    \bottomrule
  \end{tabular*}
\end{table}


\begin{table}[H]
  \centering
  \caption{Orbital Parameters.}
  \label{tab:orbital-params}
  \begin{tabular*}{\linewidth}{@{\extracolsep{\fill}}ll}
    \toprule
    Parameter & Value \\ \midrule
    Earth radius $R_{E}$     & $6,371\,\mathrm{km}$\\
    Satellite altitude $h$   & $500\,\mathrm{km}$\\
    \bottomrule
  \end{tabular*}
\end{table}


\begin{table}[H]
  \centering
  \caption{Atmospheric Parameters.}
  \label{tab:atm-params}
  \begin{tabular*}{\linewidth}{@{\extracolsep{\fill}}ll}
    \toprule
    Parameter & Value \\ \midrule
    Ground turbulence $C_{n}^{2}$   & $1.7\times10^{-14}\,\mathrm{m^{-2/3}}$\\
    Turbulence scale height         & $1\,500\,\mathrm{m}$\\
    Relative humidity               & $0.60$\\
    \bottomrule
  \end{tabular*}
\end{table}


\begin{table}[H]
  \centering
  \caption{Fundamental Constants.}
  \label{tab:const}
  \begin{tabular*}{\linewidth}{@{\extracolsep{\fill}}ll}
    \toprule
    Constant & Value \\ \midrule
    Speed of light $c$        & $3.0\times10^{5}\,\mathrm{km\,s^{-1}}$\\
    Planck constant $h$       & $6.62607015\times10^{-34}\,\mathrm{J\,s}$\\
    \bottomrule
  \end{tabular*}
\end{table}

\end{document}